\UseRawInputEncoding
\documentclass[11pt,a4paper]{article}
\pdfoutput=1

\usepackage{jheppub}

\usepackage{colortbl}
\usepackage{epsfig,multicol,bbm}
\usepackage{amsmath,amsfonts,amssymb,longtable,mathtools}
\usepackage{array}
\usepackage{graphicx}
\usepackage{subfig}
\usepackage{enumerate}
\usepackage{color}
\usepackage{placeins}
\usepackage{longtable,booktabs,caption}
\usepackage{changepage}
\usepackage{float}

\def\bea{\begin{eqnarray}}
\def\eea{\end{eqnarray}}

\def\Mp{M_{\rm Pl}}

\input epsf.sty

\usepackage[makeroom]{cancel}
  \usepackage{latexsym}
  \usepackage{epsf}
  \usepackage{amssymb}
  \usepackage{graphicx}
  \usepackage{amsmath}
  \usepackage{amsmath,amssymb,amsthm}
  \usepackage{verbatim}
  \usepackage{hyperref}
\renewcommand{\d}{\textrm{d}}

\def\ni{\noindent}
\def\bi{\begin{itemize}}
\def\ei{\end{itemize}}
\def\be{\begin{equation}}   
\def\ee{\end{equation}}
\def\ben{\begin{equation*}}
\def\een{\end{equation*}}
\def\c1{C_1}
\def\d1{D_1}
\def\ms{M_{\odot}}
\def\fpbh{f_{\rm PBH}}

\begin{document}

\title{Sharp turns in axion monodromy: primordial black holes and gravitational waves}

\author{Sukannya Bhattacharya\footnote{sukannya.bhattacharya@unipd.it}$^{a,b}$, Ivonne Zavala\footnote{e.i.zavalacarrasco@swansea.ac.uk}$^{c}$}
\affiliation{${}^a$Dipartimento di Fisica e Astronomia ``G. Galilei'', Universit\`a degli Studi di Padova, Via Marzolo 8, 35131 Padova, Italy\\
${}^b$INFN, Sezione di Padova, Via Marzolo 8, 35131 Padova, Italy\\
${}^c$Physics Department, Swansea University, SA2 8PP, UK}

\abstract{Large  turns in multifield inflation can lead to a very rich   phenomenology, but are difficult to realise in supergravity, and typically require  large field space curvatures. In this work, we present a mechanism to realise multiple sharp turns, and therefore strong non-geodesic trajectories, from   transient violations of slow-roll without the requirement of large field space curvatures in supergravity inflation.
Such  turning rates can strongly source the adiabatic fluctuations, resulting in an enhanced scalar power spectrum with resonant features and a large peak amplitude. 
If the growth of the scalar power spectrum at small scales is large enough, primordial black holes can be produced in abundance. These large  scalar fluctuations induce a  characteristic large spectrum of gravitational waves  for a wide range of frequencies, which inherits the resonant features. We illustrate this mechanism  in a supergravity model of axion monodromy, which provides 
 the first concrete model to realise such resonant features. The model  can  sustain inflation  for around  60 e-folds,
 leading to considerable production of very light primordial black holes, and large gravitational wave spectra,  which could be  detectable  by multiple upcoming gravitational wave surveys. 
For the set of parameter we consider, large oscillations occur at  all scales. This represents a challenge for the model at large scales and motivates further investigation  to reconcile this class of models with Planck data.
}

\maketitle

\section{Introduction}

Cosmological inflation is the leading mechanism to explain  the large-scale homogeneity and isotropy of  present-day universe. 
In the standard description of the inflationary epoch, the accelerated expansion is  driven by one or more scalar fields (`inflatons') which slowly roll along a nearly flat potential. 
Although single field models are in good agreement with the most recent cosmic microwave background (CMB) data~\cite{Planck:2018jri,BICEP:2021xfz},  multifield inflationary models arise  more naturally from the perspective of fundamental descriptions of gravity, such as supergravity and string theory. Moreover, single field models can have interesting growth features in the power spectra only when fine-tuned inflection points or tiny bumps or dips are incorporated in the potential.  
Exact solutions with steps in the potential inducing temporal breaking of slow-roll  were studied earlier in \cite{Starobinsky:1992ts} and used in \cite{Ivanov:1994pa} to generate PBHs and more recently  in \cite{Garcia-Bellido:2017mdw,Bhaumik:2019tvl,Mishra:2019pzq} with potentials featuring inflection points or bumps. However, multifield inflation models can have rich features and effects in the dynamics owing to the possibility of couplings (gravitational, derivative etc.) among the fields and the presence of a non-trivial field space manifold. These features leave imprints in the power spectra and thus are  of great phenomenological interest. It is one of the main goals of contemporary cosmology to understand the scale-dependence of the inflationary power spectra in view of precision data, which makes the study of such features in the multifield inflation power spectra extremely exciting and timely.  

A distinguishing feature of multifield models, compared to single field, is the existence of a new inflationary slow-roll attractor, characterised by strongly 
non-geodesic motion in  field space~\cite{RPT,Brown:2017osf,Christodoulidis:2018qdw,Garcia-Saenz:2018ifx,Garcia-Saenz:2018vqf,Bjorkmo:2019aev,Bjorkmo:2019fls,Fumagalli:2019noh,Christodoulidis:2019mkj,Christodoulidis:2019jsx}. The non-geodesicity is  measured by the dimensionless  turning rate parameter, $\omega$. Moreover, a new slow-roll parameter $\nu$ arises, associated to the change of the dimensionless turning rate per Hubble time \cite{Aragam:2020uqi,Aragam:2021scu}. From the theoretical point of view, multifield models are  attractive since  the
 $\eta_V$-problem does not arise\footnote{Here we mean that the $\eta_V$ parameter as defined in \eqref{etaV} does not have to be small in mulltifield infllation. However UV sensitivity constraints may manifest in  a different fashion. } \cite{Chakraborty:2019dfh,Aragam:2021scu},  relaxing the need for fine-tuning of the scalar potential.
The non-geodesic dynamics can lead to the enhancement of adiabatic fluctuations due to the growth of isocurvature fluctuations and curvature-isocurvature coupling. This type of deviation from geodesic motion is a novel mechanism to generate enhanced power spectrum of curvature perturbations with characteristic features. 
In particular, power spectra of the curvature perturbations with resonant features around a small-scale peak are of particular phenomenological interest~\cite{Flauger:2009ab,Flauger:2014ana,Gao:2015aba,Fumagalli:2020adf,Fumagalli:2020nvq,Braglia:2020taf,Fumagalli:2021cel,Fumagalli:2021dtd}.
Multifield models of inflation with such features can lead to abundant primordial black hole (PBH) formation due to the gravitational collapse of large  density fluctuations after horizon entry following inflation~\cite{Carr:1974nx,Carr:1975qj,Carr:2021bzv,Hawking:1971ei,Hawking:1974rv,Garcia-Bellido:1996mdl}. 
Recent studies in  supergravity  \cite{Aldabergenov:2020yok,Aldabergenov:2020bpt,Ishikawa:2021xya,Ketov:2021fww} and  field theory inflation
 \cite{Pi:2017gih,Braglia:2020eai} show that sudden broad turns arising in double inflation models can give rise to a large enhancement of the curvature perturbation, which thus  seeds PBH formation. 
However, a transient sharp turn can  greatly enhance the curvature perturbations,  seeding the generation of PBHs with enough abundance to contribute appreciably to dark matter (DM) as phenomenologically modelled in\footnote{An explicit phenomenological model realising such a peak was presented in~\cite{Anguelova:2020nzl}. } \cite{Fumagalli:2020adf,Palma:2020ejf}. 

In the second and higher orders of perturbation theory, scalar and tensor perturbations are coupled. Therefore, adiabatic perturbations source higher order tensor fluctuations~\cite{Ananda:2006af,Baumann:2007zm} (for a recent review see \cite{Domenech:2021ztg}), which are subdominant with respect to the first order tensor modes for simple slow-roll models of inflation with red-tilted adiabatic power spectra. However, an enhanced curvature power spectrum can lead to a large induced tensor power spectrum and therefore a large spectrum of induced gravitational waves (GW). Such GWs are primordial in nature, and appear as stochastic backgrounds today. With CMB surveys at large cosmological scales and ground/space-based interferometric detectors and pulsar timing arrays at smaller scales, the growing interest in using GWs as a probe for the early universe is promising to have a detailed understanding of the primordial fluctuations.
At large scales, the scalar fluctuations are  tightly constrained by CMB observations  and thus result in induced GW of tiny amplitude.
However, models where the scalar fluctuations are significantly enhanced at scales smaller than those probed by the CMB can lead to large induced GW spectra.
Thus, multifield inflation scenarios that highly  enhance the  scalar power spectrum at short scales as mentioned above,  may give rise to  potentially detectable induced GWs with interesting features 
\cite{Fumagalli:2020nvq,Braglia:2020taf,Fumagalli:2021cel,Witkowski:2021raz,Fumagalli:2021mpc,Fumagalli:2021dtd}

Due to their phenomenological richness, and potential detectability in future experiments, developing explicit models in  fundamental theories such as supergravity and string theory, realising strong non-geodesic trajectories, transient sharp turns and double inflation, opens the path to test these theories  at the highest energies through their predictions for the scalar perturbations at short scales and their associated source of  GWs. 

There has been some recent works in this direction. As mentioned above, double inflation models in supergravity generating a large scalar fluctuations to seed PBHs have been constructed in\footnote{A double D5-brane inflation model was presented in  appendix  B of \cite{Chakraborty:2019dfh}, although it presents  theoretical challenges to be consistently realised.}  \cite{Aldabergenov:2020yok,Aldabergenov:2020bpt,Ishikawa:2021xya,Ketov:2021fww}. 
On the other hand, models with  large  turns  in supergravity are rare and seem very difficult to construct, requiring large field space curvatures, as shown recently in\footnote{The only example we are aware of, is  the model in \cite{EGNO}, where  an order one dimensionless turning rate is achieved.} \cite{Aragam:2021scu}. In string theory, on the other hand, a D5-brane model with large turning rate was presented in  \cite{Chakraborty:2019dfh}. 

 An important result of  \cite{Aragam:2020uqi,Aragam:2021scu} is the realisation that sustained inflation requires the change per Hubble time of the turning rate, $\nu$ to be small (slow turns), thus introducing a new multifield slow-roll parameter. Violations of slow-roll  can therefore induce rapid turns, or strong geodesic deviations. 
This implies an obvious mechanism to induce sharp turns, or strong geodesic deviations, which we introduce in this paper. Namely,  transient large and  sharp turns are  induced through transient violations of slow-roll.  
 A natural mechanism to induce transient violations of slow-roll during inflation is through  modulations in the scalar potential. These arise naturally in axion (monodromy)  inflation due to subleading non-perturbative corrections to the axion  potential \cite{Flauger:2014ana,Kobayashi:2015aaa,CaboBizet:2016uzv,Parameswaran:2016qqq}, which can  induce a large enhancement of the density perturbations, suitable for PBH production \cite{Ozsoy:2018flq} and large induced primordial GWs. 
This mechanism also provides a novel way to `fool' supergravity, in the sense that transient violations of slow-roll induce transient strong non-geodesic motion, even when the curvature of the scalar manifold is small. 
The aim of the present work is twofold: firstly to construct a {\em multifield axion monodromy}  (AM) model\footnote{For a model of interrupted  AM  inflation - namely inflation stops ($\epsilon>1$) in stages -  with interesting phenomenology see \cite{DAmico:2020euu,DAmico:2021vka,DAmico:2021fhz}.} \cite{AM1,AM2,Flauger:2014ana} in supergravity, 
where the fields   execute strong non-geodesic motion without the requirement of a large field space curvature. The  large turns  result from transient violations of slow-roll, owing to the non-perturbative corrections in the scalar potential. 
Secondly, we show that this class of models  not only lead to a large enhancement of the adiabatic power spectra at small scales, but also provide the first concrete realisation of  the resonant features  studied recently in the literature (see e.g.~\cite{Fumagalli:2020nvq,Braglia:2020taf,Fumagalli:2021cel,Fumagalli:2021dtd}). 
These can  lead to considerable production of light PBHs and large and wide spectra of induced GWs  for a large set of fiducial parameter values. 
As we discuss, these parameter sets may not be feasible at all scales, particularly, it is difficult to reach CMB consistent power spectrum at large scales while such oscillations are present.  Nevertheless, the phenomenology of the mechanism and the model is very rich and interesting, as we demonstrate with various examples, while leaving for a future work a careful exploration of this class of models and its viable parameter space.

The paper is organised as follows. In section~\ref{Sec1} we introduce our notation, briefly review multifield inflation and introduce our mechanism to generate large turns from slow-roll violations. In section~\ref{Sec2} we introduce the supergravity axion monodromy model and discuss its background evolution with a sample parameter set. In section~\ref{Sec3} we present the cosmological perturbations and numerically evaluate the adiabatic power spectra for some fiducial set of parameters. 
We discuss the growth of the power spectra with resonant oscillations at small scales. 
In a dedicated subsection, we discuss the power spectrum at CMB scales in more detail.
In section~\ref{PBHPGW}, we discuss the phenomenological implications in terms of production of PBH and large GW spectra due to the enhanced adiabatic fluctuations in our model. We  discuss our results and conclude in section~\ref{Sec4}. 

\section{ Multifield inflation and large turns}\label{Sec1}

We start by briefly reviewing slow-roll multifled inflation following ref.~\cite{Aragam:2021scu}, focusing on the  two  field case.  
The starting point is  the following lagrangian
\be\label{4Daction}
S= \int{d^4x \sqrt{-\tt g} \left[\Mp^2 \frac{R}{2}  - \frac{g_{ab}}{2} \partial_\mu\phi^a \partial^\mu\phi^b - V(\phi^a)\right]} \,,
\ee
where $g_{ab}$ is the field space metric and ${\tt g} = {\rm det}\,{\tt g}_{\mu\nu}$, where ${\tt g}_{\mu\nu}$ is the FRW metric. $\Mp$ is the reduced Planck mass.
The equations of motion in an FRW spacetime, projected along the tangent and normal (i.e.~adiabatic and entropic) directions to the inflationary trajectory, are given by 
\bea
&&H^2 = \frac{1}{3\Mp^2} \left(\frac{\dot \varphi^2}{2}  + V(\phi^a)\right) \,, \label{H} \\
&& \ddot\varphi + 3H\dot\varphi + V_T =0 \,, \label{varphiT}\\
&& D_t T^a \equiv -\Omega N^a \label{Omega1}\,,
\label{eq:EOMs}
\eea
where $T^a = \frac{\dot\phi^a}{\dot\varphi}\,, $ with $T^aT_a =1$ is the tangent (adiabatic)  and $N^a$ with $N^aT_a=0$, $N^aN_a=1$ is the normal (entropic) directions along the trajectory. The velocity is given by
\be\label{varphi}
\dot\varphi^2 \equiv g_{ab} \dot \phi^a\dot\phi^b\,,
\ee
and we introduced the  {\em turning rate} parameter $\Omega$ and we    define the dimensionless turning rate as 
\be\label{omega}
\omega\equiv \frac{\Omega}{H} \equiv \frac{V_N}{H\dot\varphi}\,,
\ee 
which measures the departures from the geodesic trajectory for $\omega \gtrsim1$. Finally, the directional derivative is given by 
 \be\label{Dt}
 D_tT^a = \dot T^a + \Gamma^{a}_{bc} T^b \dot \phi^c \,,
 \ee
where the Christoffel symbols  are computed using the scalar manifold metric $g_{ab}$, and $V_T = V_aT^a$, $V_N = V_aN^a$ with $V_a$ the derivative w.r.t the scalar field $\phi^a$. 

Using the equations of motion, we can  write  the projections of the Hessian elements along the tangent vector as  \cite{Achucarro:2010da,Hetz:2016ics,Christodoulidis:2018qdw,Chakraborty:2019dfh}: 
\be\label{VTT1}
\frac{V_{TT}}{3H^2} =  \frac{\Omega^2}{3H^2} + 2\,\epsilon -\frac{\eta}{2} -
\frac{\xi_\varphi}{3}   \,,
\ee
as well as the projection along $T $ and $N$ as \cite{Aragam:2021scu}:
\be\label{VTN1}
\frac{V_{TN}}{3H^2} = \omega\left( 1-\epsilon+\frac\eta3+\frac\nu3 \right).
\ee
In these equations  we  introduced the {\em slow-roll} parameters:
\begin{subequations}
\begin{align}
\label{epsilonH}
\epsilon&\equiv -\frac{\dot H}{H^2} = \frac{\dot\varphi^2}{2M^2_{Pl}H^2} \,,\\
\eta &\equiv \frac{\dot\epsilon}{H\epsilon}=2(\delta_\varphi+\epsilon)\,,\label{eta} \\
\delta_\varphi&\equiv \frac{\ddot{\varphi}}{H \dot{\varphi}} \,,\label{etaphi} \\
\xi_\varphi &\equiv \frac{\dddot\varphi}{H^2 \dot\varphi}\,,    \\
\nu&\equiv \frac{\dot\omega}{H\omega}\,. \label{nu}
\end{align}
\end{subequations}

\smallskip

\ni Note that the expressions \eqref{VTT1}, \eqref{VTN1}  are exact, as we have not made use of any slow-roll approximations. 
 On the other hand, $V_{NN}$ depends on the inflationary trajectory in a model-dependent manner. 

\subsection{Slow-roll inflation}

The slow-roll conditions  require the  slow-roll parameters $\epsilon, \eta, \delta_\varphi $, defined above, to be much smaller than one to guarantee long lasting slow-roll inflation, that is,  $\epsilon, \eta, \delta_\varphi, \xi_\varphi \ll 1$. 
These conditions imply 
\bea
&&H^2 \simeq \frac{V}{3M_{Pl}} \,, \\
&& 3 H\dot \varphi + V_T \simeq 0 \,, 
\eea
and thus that the tangent projection of the derivative of the potential is small, that is:
\be\label{epT}
\epsilon_T \equiv \frac{M_{Pl}^2}{2} \left(\frac{V_T}{V}\right)^2 \ll1  \,.
\ee
On the other hand, the normal projection $V_N$ does not need to be small, and it is related to the turning rate by eq.~\eqref{omega}. 
Additionally, from \eqref{VTT1} we see that during slow-roll,
\be\label{VTTsr}
\frac{V_{TT }}{3H^2} \sim \frac{\Omega^2}{3H^2}\,,
\ee
while from \eqref{VTN1} we observe that, barring cancellations, $\eta\ll1$ (equivalently $\delta_\varphi \ll 1$), implies that
\be\label{VTN_slowroll}
   \frac{V_{TN}}{3H^2}\sim \frac{ \Omega}{H} \,, \qquad {\rm and} \qquad 
     \nu \ll 1\,.
\ee
Hence, we see that $\nu$ behaves as a new slow-roll parameter in multifield inflation: the turning rate is guaranteed to be slowly varying during slow-roll \cite{Aragam:2020uqi,Aragam:2021scu}.
As discussed in~\cite{Aragam:2021scu}, the slow-roll conditions above do not require small eigenvalues of the Hessian. That is, the $\eta_V$ parameter:
\be\label{etaV}
\eta_V \equiv \Mp^2 \left|{\rm min\,\,\, eigenvalue}  \left(\frac{ \nabla^a \nabla_b V}{V}\right)\right|, 
\ee
does not need to be small  in multifield inflation and  can indeed be much  larger than one\footnote{Interestingly, as pointed out in \cite{Garg:2018reu}, $\eta_V\gtrsim {\cal O}(1)$ might be required in consistent theories of  quantum gravity. From this point of view, multifield inflation models with $\eta_V\gtrsim1$  would be consistent with this constraint.}, as in the examples discussed in \cite{Chakraborty:2019dfh,Aragam:2021scu}.

 \subsection{Sharp turns from transient slow-roll violations}
 
  The discussion above implies that if the slow-roll condition is obeyed exactly with $\epsilon, \eta, \delta_\varphi \ll 1$, then the field moves in the manifold with slow-turns,  $\nu\ll1$. This hints at the possibility of obtaining large turning rates, and therefore strong deviations from a geodesic trajectory, if one or more of the slow-roll conditions are violated, while still maintaining a long-lasting inflationary paradigm that is consistent with current observations.
Indeed,  if the potential has intrinsic features which give rise to transient violations of the slow-roll condition with $\eta\gtrsim 1$, it will  generate transient violations of slow-turn (leading to sharp turns),  or strong geodesic deviations, with $\omega \gtrsim 1$ and $\nu \gtrsim 1$. This interesting effect arises naturally  in multifield axion inflation in field theory and supergravity and we  study this mechanism in what follows. 

The axion monodromy model that we consider below is particularly interesting  since it leads to growth of perturbations by exploiting the relation between two parameters, $\eta$ and $\omega$, via Eqs.~\eqref{VTT1} and~\eqref{VTN1}. 
The transient large values of $\eta$, and therefore of $\delta _{\phi}$,  repeatedly induce kicks in the the turning rate, which becomes sharp and large\footnote{This scenario is a combination of more complicated versions of the two types of features presented in~\cite{Boutivas:2022qtl}, as we discuss later.}.

\section{Multifield axion monodromy in supergravity}\label{Sec2}

We now  construct a supergravity  axion monodromy model, which is the two field realisation of the single field model  introduced  in \cite{Ozsoy:2018flq}.

The scalar potential in supergravity is constructed from the  K\"ahler potential, $K(\Phi, \bar\Phi)$, which is a real function of the superfields $\Phi, \bar\Phi$, whose scalar component is the complex field; and the holomorphic superpotential, $W(\Phi)$, as 
\be
V= e^{K/M_{\rm Pl}^2} \left(K^{i\bar{\jmath}} D_iW \overline{D_jW} - 3 |W|^2 M_{\rm Pl}^{-2}\right)\,,
\ee
 where $D_i W = W_i + (K_i/\Mp) W$, with $W_i \equiv \frac{\partial W}{\partial\Phi_i}$ and $K_{i\bar{\jmath}}$ is  the K\"ahler metric, which when passing to real coordinates, can be identified with the field space metric introduced in \eqref{4Daction} as $2 K_{i\bar \jmath} = g_{ab}$.
 
We  use the approach in \cite{Kawasaki:2000yn,KLR,Ferrara:2014kva} and introduce two ``orthogonal" chiral superfields  \cite{Roest:2013aoa}, the goldstino, $S$, and inflaton superfield, $\Phi$, where we denote the scalar components of these superfields with the same letter. We then eliminate the sgoldstino, $S$ by either introducing  suitable K\"ahler potential  corrections to stabilise it \cite{KLR}, or simply by introducing a nilpotent condition on it, $S^2=0$ \cite{Ferrara:2014kva}. 

The K\"ahler and superpotentials  for the  axion monodromy model are  given by 
 \bea\label{eq:KW}
&&\Mp^{-2}\,K = - \alpha \log[(\Phi + \bar \Phi)/\Mp -\beta S\bar S/\Mp^2] \,, \\
&&W = S(M\Phi + i \lambda e^{- b\Phi} )\,.
\eea
The K\"ahler potential is independent of the imaginary part of the inflaton superfield, Im$(\Phi)=\theta$,   the {\em axion}, while it depends only on the  the real part, Re$(\Phi)=\rho$, the {\em saxion}. The shift symmetry of the axion is broken by the non-perturbative term in the superpotential, as well as at  tree-level by the linear term. In a string theory set-up, the inflaton could be identified with a complex structure modulus, with the fluxes breaking the shift symmetry at tree-level (see for example  \cite{Kobayashi:2015aaa,CaboBizet:2016uzv}). 

We choose $\alpha =1$ as  in \cite{Ozsoy:2018flq} for which the K\"ahler metric is $K_{\Phi\bar\Phi} = 1/(\Phi+\bar\Phi)^2$ and thus the field space  curvature is ${\mathbb R} = -4$. The scalar potential becomes: 
\be\label{eq:V}
V = \frac{M^2 }{\beta} \left(\rho^2 + \theta^2 + \frac{2\lambda}{M} \,e^{-b \rho} \left[\theta\,  \cos{(b \,\theta)}+  \rho\,  \sin{(b\, \theta)} + \frac{\lambda}{2M} \,e^{-b \rho}\right] \right)\,,
\ee
where we used that  $\Phi= \rho+ i\theta$.
In contrast to the case when the saxion $\rho$ has been stabilised \cite{Ozsoy:2018flq}, the modulations are now saxion dependent, and  damped by the exponential terms. 
The structure of the modulations along the axion depends on the parameters $M,\lambda, b$, and the value of the saxion, and it is encoded in the condition:
\be
x\left[c\, e^y - \sin x \right] = - \cos x (1+y) \,,
\label{cxy}
\ee
where we defined $c\equiv \frac{M}{\lambda \,b}$,  $x \equiv b\, \theta$, $y\equiv b \rho $. For $y=0$, the potential has an infinite number of stationary points if $c <1$, while  for larger values of $c$ there will be a finite number of  stationary points. Once we introduce $y$, this behaviour  depends on the saxion's value and the modulations of the axion are  strongly damped by the saxion field values (we consider only positive saxion values). 
When the axion and saxion are displaced from their minima, they will evolve traversing the modulations in the potential \eqref{eq:V}. The parameter $1/b$ acts as a ``decay constant" for the axion at a fixed value of the saxion, while we can define an {\em instantaneous decay constant} as $f_{\rm inst} =\sqrt{g_{\theta\theta}}/b$,  \cite{Chakraborty:2019dfh}. We consider sub-Planckian values of $1/b$  in agreement with recent quantum gravity constraints  on axions \cite{WGC}. We can then fix the value of $\lambda/M$ to determine the size of the modulations as the axion-saxion system evolves. The parameters $M$ and $\beta$ fix the amplitude of  the power spectrum. We do not make a thorough search in the parameter space in the present work, but make a selection of  parameters that allow us to demonstrate the following aspects: 
\begin{enumerate}[{\bf i.}]
\item Transient large turning rates can be generated in supergravity with small field space curvature - fooling supergravity - through  transient violations of slow-roll, albeit sustaining enough inflation. Moreover, the minimal eigenvalue of the Hessian is large (in Hubble units) and tachyonic, as conjectured in~\cite{Aragam:2021scu}.

\item   Multifield axion monodromy in supergravity naturally gives rise to transient violations of slow-roll and thus transient large turns, due to modulations of the axion potential from (leading and subleading) non-perturbative effects. These give rise to distinctive resonant features in the power spectrum, providing the first concrete model with such  
features in multifield inflation,  and  interesting phenomenology as we discuss in section~\ref{PBHPGW}.
\end{enumerate}

\subsection{Transient large turns  in supergravity }

\begin{figure}[t]
\center{
\includegraphics[width=0.6\textwidth]{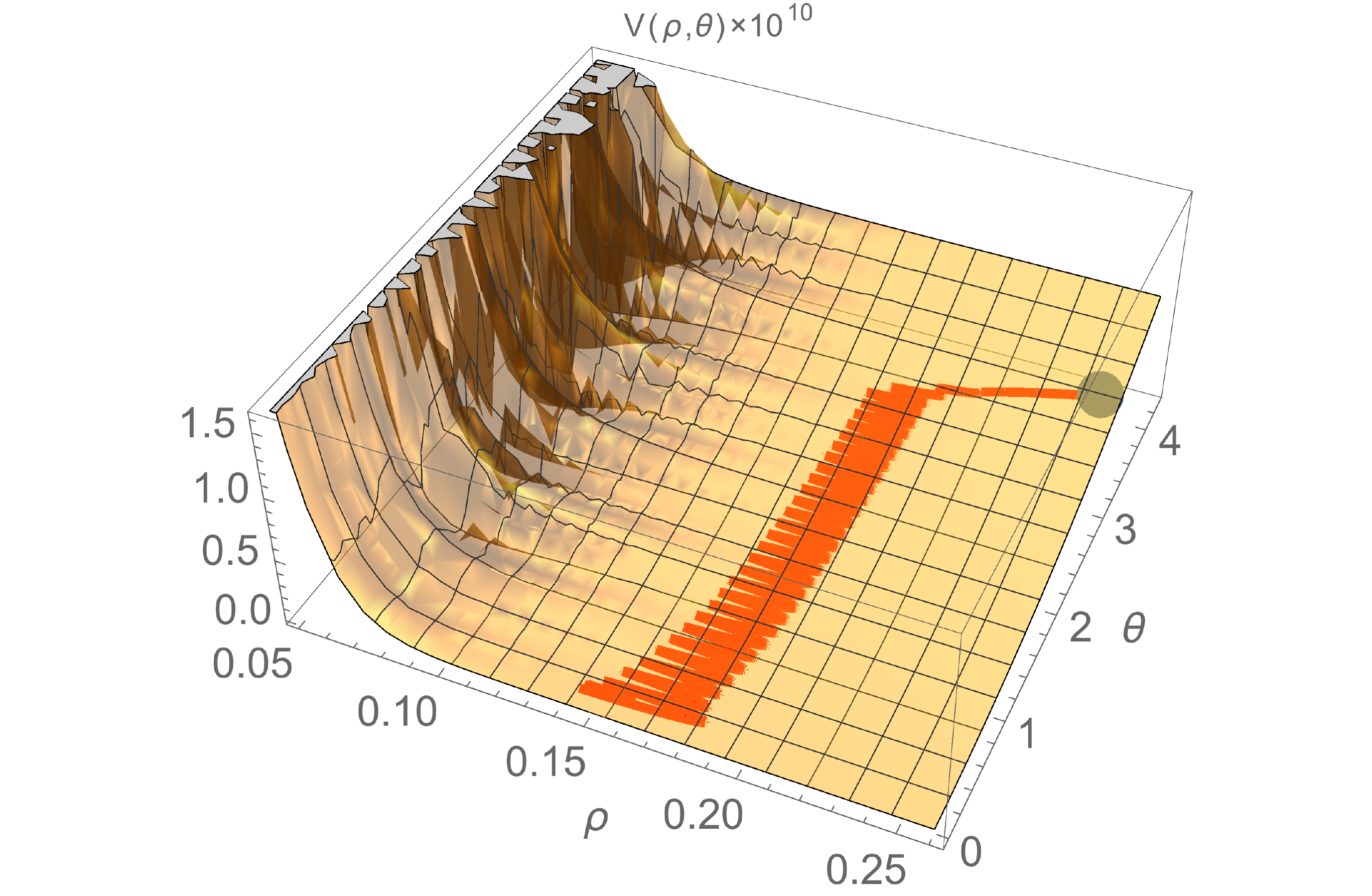}}
\caption{Inflationary trajectory of axion-saxion system as they move in the scalar potential \eqref{eq:V} for the parameter values in Table~\ref{tab:1}.}
\label{fig:V}
\end{figure}

\begin{figure}[h]
\center{
\includegraphics[width=0.48\textwidth]{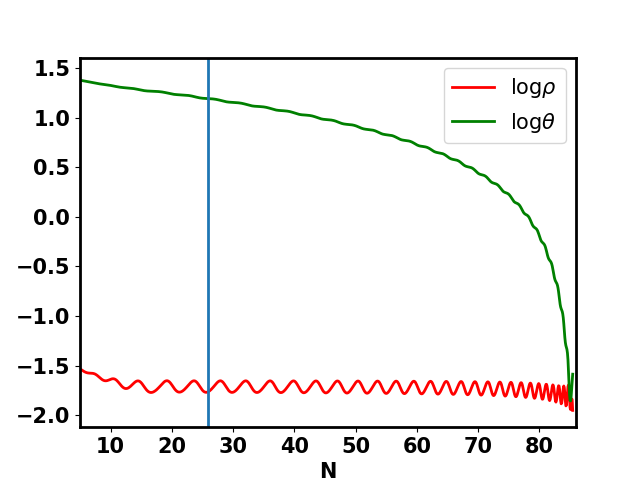}
\includegraphics[width=0.48\textwidth]{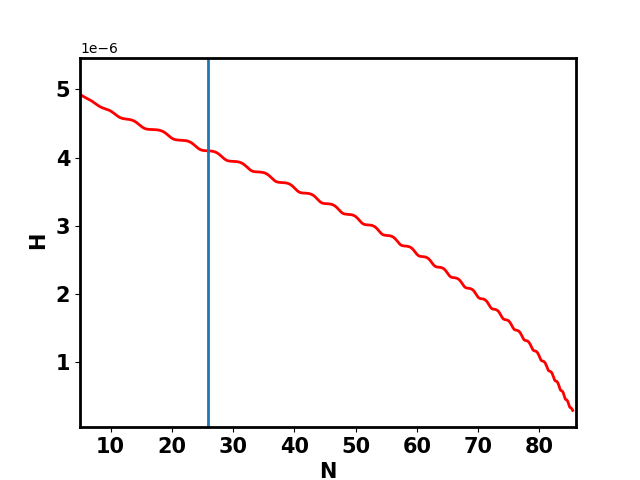}}
\caption{The field evolution and the Hubble parameter  as a function of the number of e-folds $N$ during inflation for the set  of parameters in Table \ref{tab:1}. The blue vertical line corresponds to $N_{\rm pivot}$.}
\label{fig:fieldomega}
\end{figure}

As we discussed in section \ref{Sec1}, sustained slow-roll implies $\nu\ll 1$ (see eq.~\eqref{VTN_slowroll}). Thus, a violation of the slow-roll condition, $\eta\ll1$, implies $\nu\gtrsim1$, which means that the turning rate is changing rapidly. When the saxion is kept fixed at its minimum, the axion  evolves  as discussed in~\cite{Ozsoy:2018flq}. 
Namely, for $c<1$ in Eq.~\eqref{cxy}, there are several minima and the axion may get trapped in one of them. On the other hand, for suitable values of $c$, the axion potential develops an inflection point at which there is a transient violation of slow-roll  suitable for sufficient enhancement of the adiabatic power spectrum for abundant PBHs production~\cite{Ozsoy:2018flq}.
Some amount of fine-tuning is required to keep the large scale amplitude of the power spectrum $~\mathcal{O}(10^{-9})$~\cite{Planck:2018jri}. In the  multifield case, the axion-saxion system is non-trivially coupled, and thus, when the axion is displaced from its minimum,  the saxion is also displaced slightly, and will follow its adiabatic minimum as the axion evolves (see Fig.~\ref{fig:V}). The modulations of the potential at the minimum generically have several stationary points. However, away from the minimum the modulations are suppressed by the saxion, giving  very gentle plateaus and cliffs. These will generically give rise to transient violations of slow-roll as the axion moves along these plateaus and cliffs repeatedly before reaching its minimum. This motion causes transient violations of slow-roll, $\eta\gtrsim1$, which generate transient violations of the slow-turn condition $\nu\gtrsim1$,  inducing transient large non-geodesic trajectories (see Fig.~\ref{fig:fieldomega}). 

\begin{figure}[h]
\center{
\includegraphics[width=0.49\textwidth]{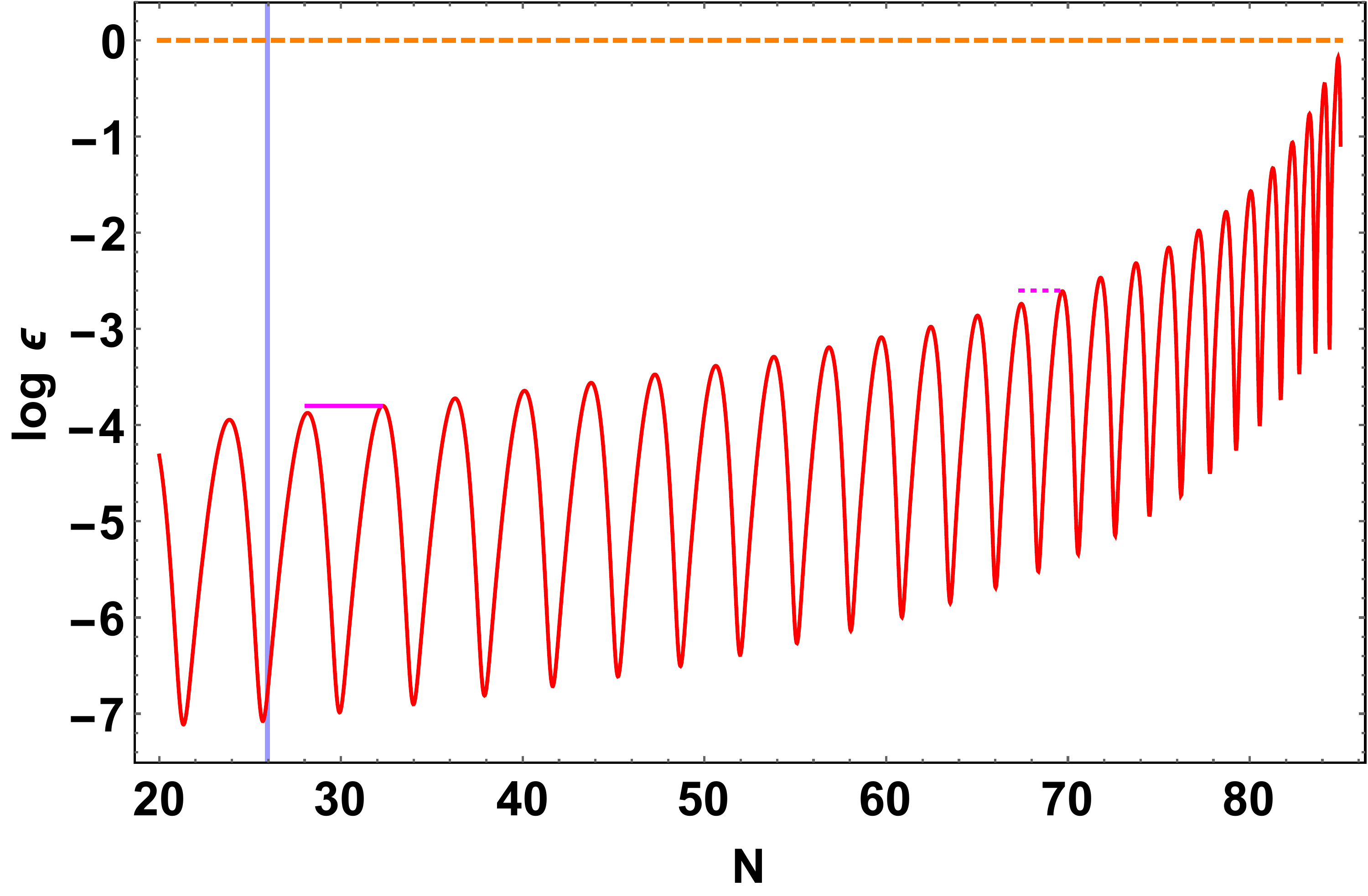}
\includegraphics[width=0.49\textwidth]{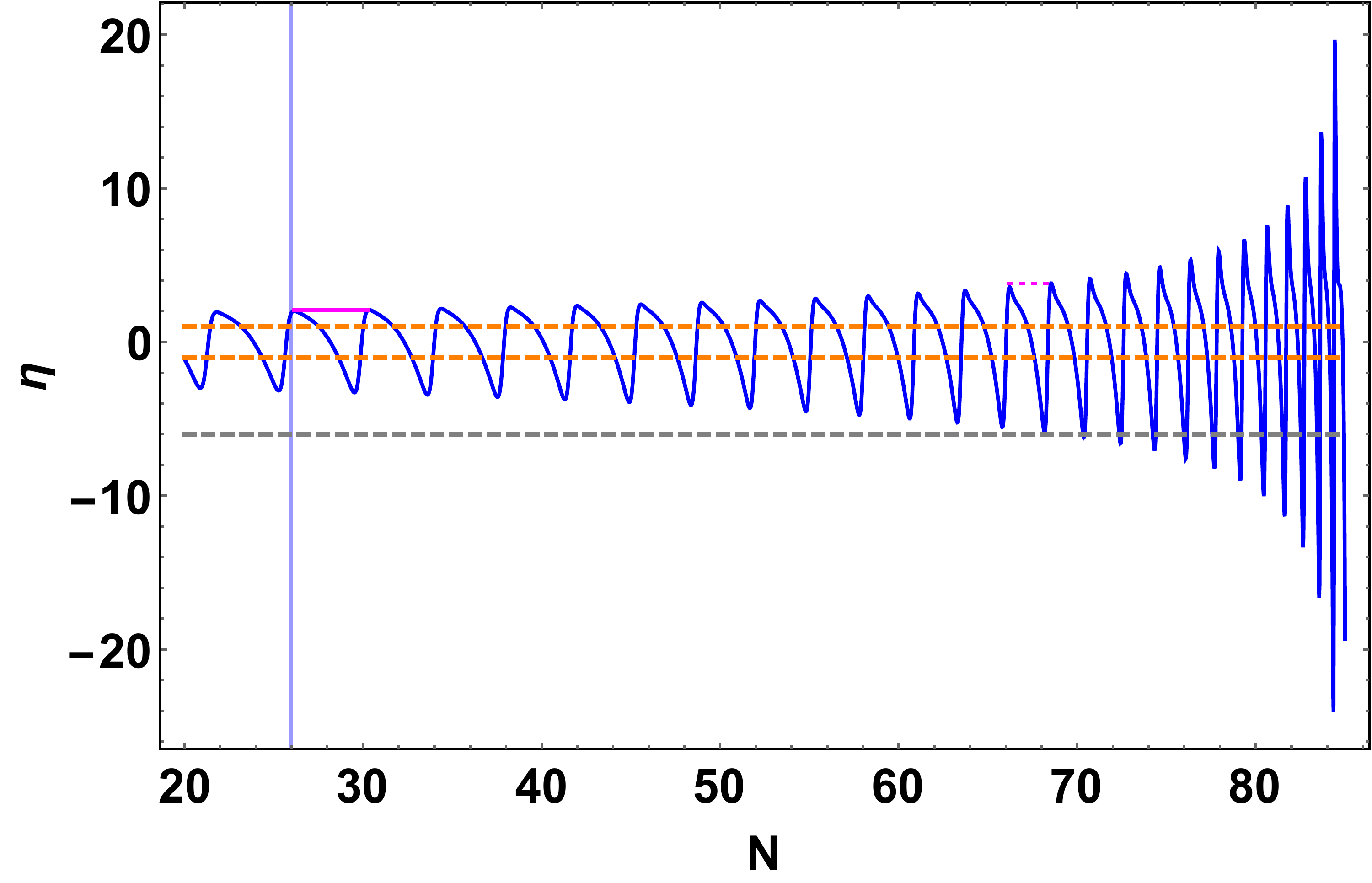}
\includegraphics[width=0.49\textwidth]{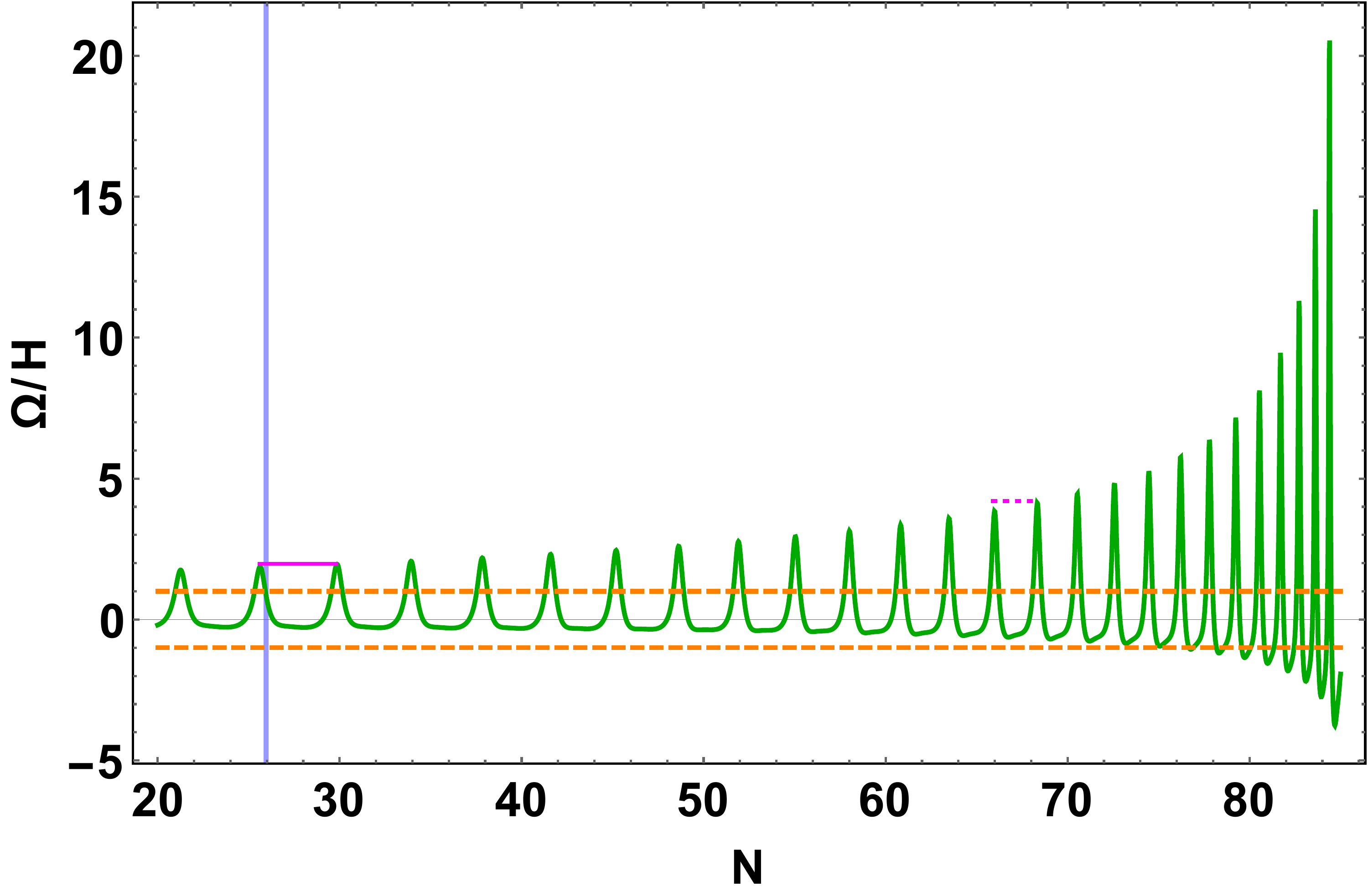}
}
\caption{Variation of the slow-roll parameters for $\epsilon$,  $\eta$ and the dimensionless turning rate  $\omega = \Omega /H$  as functions of $N$. In the plots of $\eta$ and $\omega $, the dashed lines signify the boundaries $1, -1$. The gray dashed horizontal line signifies the ultra slow-roll limit $\eta =-6$ in the $\eta$-plot. The blue vertical lines correspond to $N_{\rm pivot}$. All the curves here correspond to the evolution for the parameters' set  in Table~\ref{tab:1}. Magenta horizontal lines in each plot denote the periodicity of the oscillations.}
\label{bkgd_fig1}
\end{figure}

We choose a generic set of parameters given in Table~\ref{tab:1} to illustrate  the  background evolution.  As mentioned above, we choose $1/b$ to be sub-Planckian,  $b=50\Mp^{-1}$, and then fix $\lambda/M$ aiming at a large enhancement of the power spectrum. The normalisation of the power spectrum at CMB scales is fixed by $M^2/{\beta}$. We also choose suitable initial conditions to ensure enough e-folds of inflation \footnote{Note that no substantial fine tuning of the parameters and/or initial conditions is required, as it is the case for single field~\cite{Ozsoy:2018flq}.}, 
although the adiabatic power spectra are in tension with CMB bounds for an oscillatory  spectrum (see next section).
 The background cosmological evolution is shown in Figures~\ref{fig:V}-\ref{bkgd_fig1}.
In Fig.~\ref{fig:V} we show the evolution of the fields as they move in the potential~\eqref{eq:V}. Let us  stress that this potential does not require  any special properties to feature several turns in  field space, generating transient strong non-geodesic trajectories. It  arises naturally from the  multifield axion monodromy  potential described by~\eqref{eq:KW}.
In Figure~\ref{fig:fieldomega}, the excursion of the two fields $\rho$ and $\theta$ and the evolution of the Hubble parameter as functions of the number of e-folds are shown.
In  figure  \ref{bkgd_fig1} we   show the evolution of the slow-roll parameters as functions of the number of e-folds, which demonstrate 
 the mechanism at work to generate transient non-geodesic motion. Namely transient violations of slow-roll ($\eta\gtrsim 1$) induce brief violations of slow-turns ($\nu\gtrsim 1$), thus causing short-term large turns: $\omega\gtrsim1$. 

\begin{table}[!h]
\centering
\begin{tabular}{|c|c|c|c|c|c|}
\hline
 \cellcolor[gray]{0.9}$\beta$ &\cellcolor[gray]{0.9} $b$ & \cellcolor[gray]{0.9}$M$ & \cellcolor[gray]{0.9}$\lambda / M$ & \cellcolor[gray]{0.9}$\rho _{\rm ini}$ & \cellcolor[gray]{0.9}$\theta _{\rm ini}$
\\
\hline
 1 & 50 & $2.15 \times 10^{-6}$ & 80 & 0.245 & 4.20
\\
\hline
\end{tabular}
\caption{Parameter values in Planck units (except $\beta$, which is dimensionless).
}
\label{tab:1}
\end{table}

For the parameters in Table~\ref{tab:1}, the value of $c\equiv \frac{M}{\lambda b}=2.5\times 10^{-4} \ll 1$ and therefore the potential has infinitely many stationary points at $y=b\rho =0$. At the non-vanishing value of $\rho=\rho_{\rm min}\simeq0.13$, the potential for $\theta$ has several stationary points, while during inflation, these modulations are exponentially damped, allowing for inflation to occur. These  non-trivial oscillations along the whole evolution, will be inherited to the cosmological perturbations that we 
discuss in the next section. 
In particular the ``periodicity" of the modulations due to the potential are directly inherited into the slow-roll parameters and the turning rate, which are shown with the magenta horizontal lines in Fig.~\ref{bkgd_fig1}. These oscillations influence the oscillations in the power spectra as we  see in the next section. 

\section{Cosmological perturbations}\label{Sec3}

 The linear perturbations in multifield inflation can be neatly described decomposing them in terms of  adiabatic and entropic  modes, $Q_T$, $Q_N$, respectively, defined  as  the  projections  of the field fluctuations $Q^a$ in spatially flat gauge \cite{SS,GWBM,GNvT,LRP}. 
The equations of motion describing the dynamics of the primordial linear perturbations about the inflationary background  are given by  \cite{SS,GNvT,LRP}:
\bea
&&\ddot Q_T + 3H\dot Q_T + \left(\frac{k^2}{a^2} +m_T^2  \right) Q_T = 
\left(2\omega H Q_N\right)^{\Large\dot{}}-\left(\frac{\dot H}{H} + \frac{V_T}{\dot\varphi} \right) 2\,\omega H Q_N\,, \label{QT} \\
&& \ddot Q_N + 3H\dot Q_N + \left(\frac{k^2}{a^2} +m_N^2  \right) Q_N =- 2\,\omega 
\dot \varphi \dot {\cal R}   \label{QN}
\eea
where $Q_T=T_i Q^i $, $Q_N=N_i Q^i$, $Q^i$ are the field fluctuations in spatially flat gauge, ${\cal R}$ is the comoving curvature perturbation directly proportional  to the adiabatic fluctuation through: 
\be
{\cal R} = \frac{H}{\dot \varphi} Q_T\,.
\ee
The adiabatic mass $m_T$ is given by
\be\label{amass}
\frac{m^2_T}{H^2}  \equiv -\frac{3}{2}\eta- \frac{1}{4}\eta^2 -\frac{1}{2} \epsilon\eta-\frac{1}{2}\frac{\dot\eta}{H} \,,
\ee
while  the entropic mass, $m_N$, is given by 
\be\label{emass}
\frac{m_N^2}{H^2} = \frac{V_{NN}}{H^2} + \Mp^2\, \epsilon  \,{\mathbb R} - 
 \omega^2
\,,
\ee
At superhorizon scales, 
\be
 \dot{\cal R} \simeq 2\,\omega \frac{H^2}{\dot\varphi} Q_N\,,
\ee
and \eqref{QN} becomes
\be
 \ddot Q_N + 3H\dot Q_N + \left(m_N^2 +4H^2\omega^2 \right) Q_N \approx 0\,,
\ee
where it is useful to define an effective entropy mass as
$$M_{eff}^2 \equiv m_N^2 + 4H^2\omega^2 = V_{NN} + \Mp^2\, \epsilon  \,{\mathbb R}H^2 + 3H^2\omega^2.$$ 
\begin{figure}[b]
\begin{center}
\includegraphics[width=0.7\textwidth]{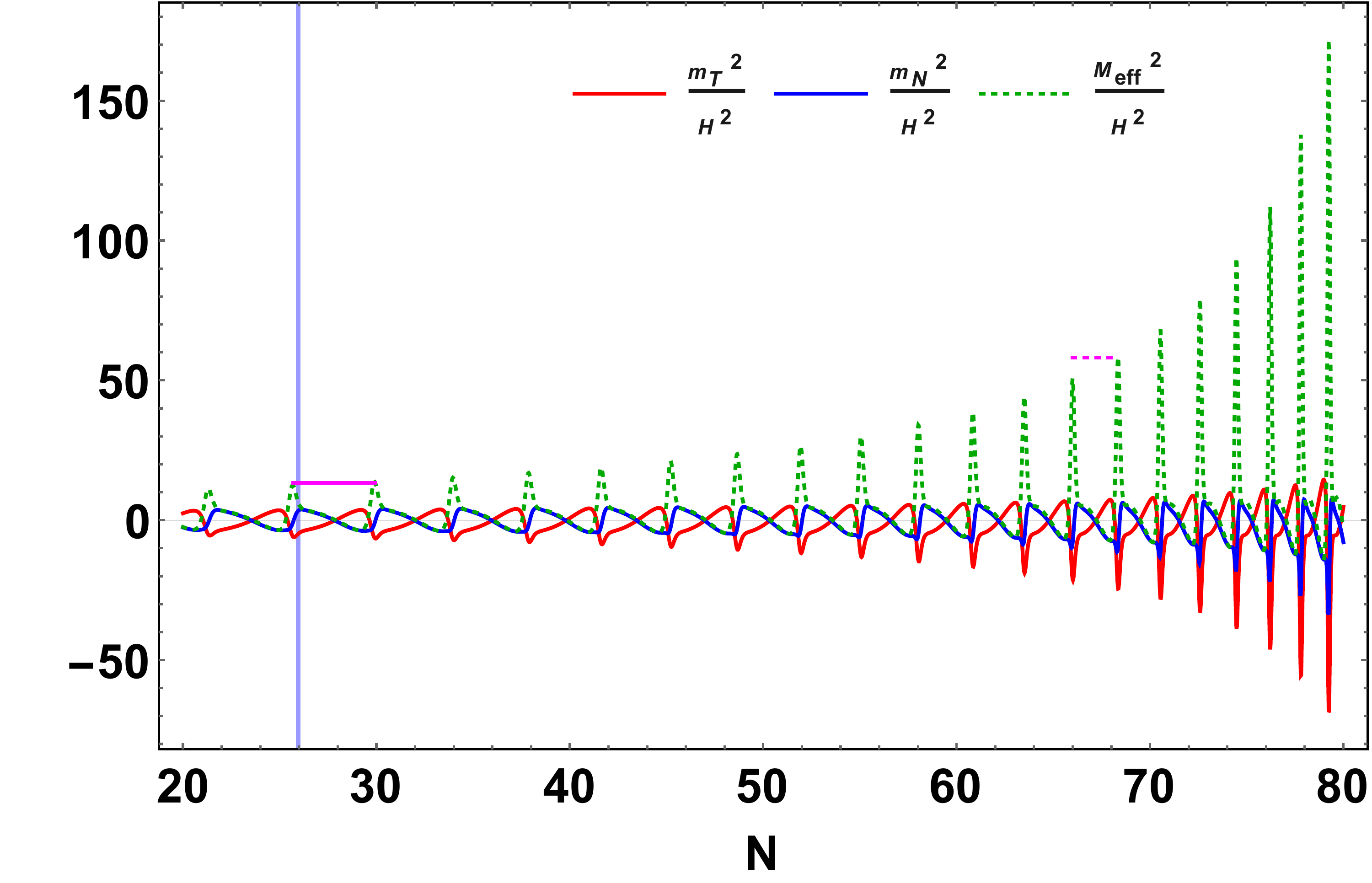}
\caption{Evolution of the adiabatic and entropic masses with respect to Hubble: $\frac{m_T^2}{H^2}$ and $\frac{m_N^2}{H^2}$ in Eqs~\eqref{amass} and~\eqref{emass} respectively with the number of e-folds. The effective mass is also shown as $\frac{M_{\rm eff}^2}{H^2}$.
All the curves here are for the example in Table~\ref{tab:1}. Magenta horizontal lines denote the periodicity of oscillations in $M_{\rm eff}^2$.}
\label{fig:adenmass}
\end{center}
\end{figure}
The dynamics of the linear perturbations, and therefore cosmological predictions  depend on 
the  hierarchies of the adiabatic and entropic modes' masses relative to each other, which in turn depend on the slow-roll parameters, the turning rate $\omega$ and its variation $\nu$, the curvature of the scalar manifold  ${\mathbb R} $, and  $V_{NN}$. 
For example, if   ${\mathbb R} $ is  negative and large, it may trigger geometric destabilisation of the entropy modes as discussed in \cite{RPT}.
Notice that the adiabatic mode will  be light (w.r.t.~$H$) as long as slow-roll is satisfied (see eq.~\eqref{amass}). 

\subsection*{Perturbations in supergravity axion monodromy}

In the axion monodromy model described by the potential  Eq.~\eqref{eq:V}, the cosine and sine terms  induce large oscillations in the background parameters as we discussed above. The transient violations of the slow-roll condition with $\eta \gtrsim1$ lead to multiple points of strong deviation from the geodesic motion with $\nu \gtrsim1$ and thus $\omega \gtrsim1$. 
Note that $\epsilon\ll1$ always in Fig.~\ref{bkgd_fig1}, that is, inflation does not stop intermediately. 
 These details enter the scalar fluctuations through the Hubble parameter $H$ and its derivatives, the masses $m_T^2$ and $m_N^2$, $\omega$ and $\nu$ (see Eqs.~\eqref{QT}-\eqref{QN}). The background oscillations are imprinted in the Hubble parameter (right panel of Fig.~\ref{fig:fieldomega}), the slow-roll parameters $\epsilon$, $\eta$  and in the turn rate $\omega$ (Fig.~\ref{bkgd_fig1}). These lead to oscillatory features in the adiabatic and entropic masses (Fig.~\ref{fig:adenmass}), 
with a  ``decreasing periodicity" inherited from the scalar potential (shown with magenta lines in the figure). 
\begin{figure}[h]
\center{
\includegraphics[width=0.8\textwidth]{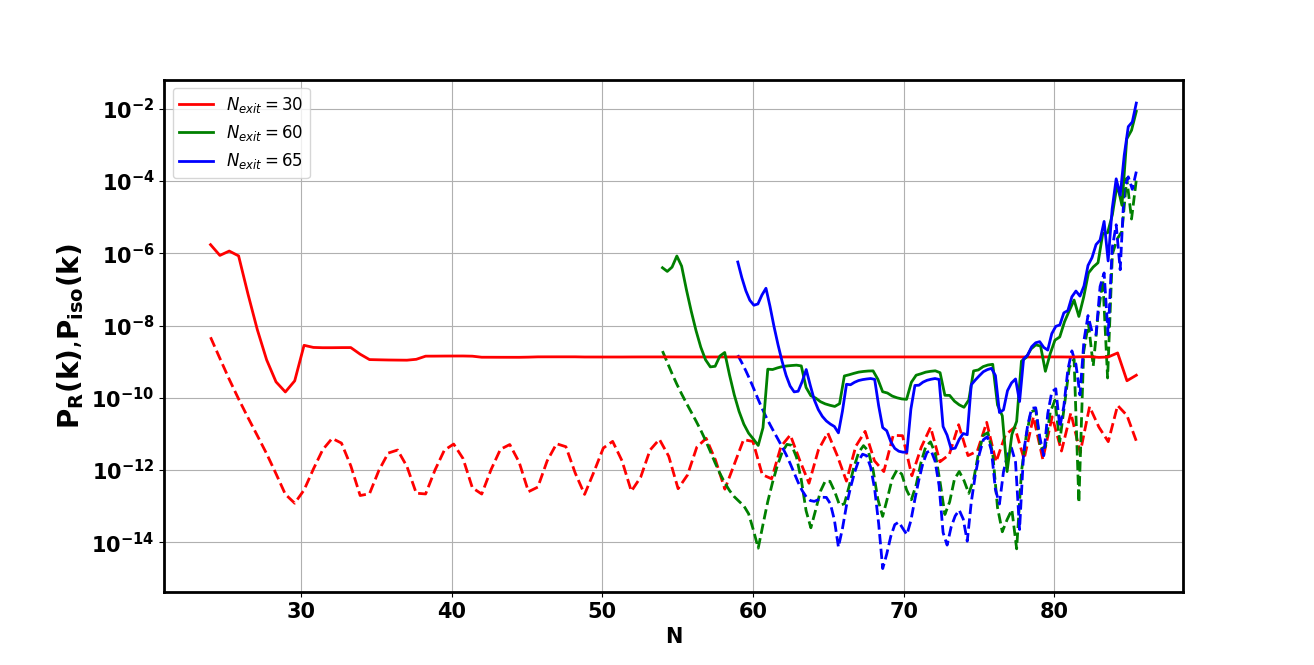}
}
\caption{Evolution of the curvature and the isocurvature power spectra for three different modes which exit the horizon at $N=30$, $N=60$ and $N=65$ respectively. All the curves here are for the example in Table~\ref{tab:1}. Plotted using \texttt{PyTransport}. }
\label{fig:psk1k2}
\end{figure}
 The oscillatory effects from the background and multiple sharp turns in field space source the scalar fluctuations in a cumulative manner.

The resulting turns in field space are numerous and sharp for all the parameter space values where the large scale power spectrum is as close as possible to the  CMB, with an amplitude $\sim 10^{-9}$ and red-tilt of the oscillatory envelope, while it grows substantially at small scales. 
The relation of the number and position of the turns (i.e.~of slow-roll violations) with the parameters is rather involved, given its non-trivial dependence on the evolution of the saxion as  encoded in eq.~\eqref{cxy}.
A piecewise analysis for the contributions to the growth of adiabatic fluctuations is thus difficult in the present scenario due to the presence of inherent background oscillations and multiple rapid turns in the field space.
Nevertheless, we can see a clear pattern in the modulations. Namely, there is a ``decreasing periodicity" as the fields move traversing the potential softened modulations at the beginning, which become stronger, as the saxion decreases.  For the parameter choice in Table \ref{tab:1}, we observed that the ``periodicity" of the parameters inherited from the potential starts at about $N\sim 4$ at the pivot scale, decreasing to $N\sim 2$ towards the end of inflation. 
An attempt to understand this analytically  would be through means of simpler modelled slow-roll violations as for example presented in~\cite{Boutivas:2022qtl}: large turning rates lead $m_{\rm N}^2$ to large negative values (see Fig.~\ref{fig:adenmass}), which make the isocurvature fluctuations  grow (see Eq.~\eqref{QN}), and in turn, act as a source for the adiabatic perturbations (see Eq.~\eqref{QT}). This results in the growth of curvature fluctuations and they contribute as a source for the isocurvature fluctuations.
As a result, the modulations that can be seen in the evolution of  a single scalar mode $Q_T^k$ are a combined effect of oscillations in background parameters and interference of characteristic oscillations due to sharp turns in field space. 

For the set of parameters in Table \ref{tab:1} with $N_{\rm pivot}\simeq 26$, the evolution of the adiabatic and isocurvature power spectra for three distinct modes $k_1$, $k_2$ and $k_3$ is shown  in Fig.~\ref{fig:psk1k2}, which  exit the horizon at $N=30$, $N=60$ and $N=65$ respectively. For $k_1$, the curvature fluctuations grow very little after horizon exit due to a few number of field space turns executed before $N=30$ and become frozen soon after horizon exit (within ~$5$ e-folds). But, for $k_2$ and $k_3$, superhorizon evolution is greatly affected by the turns in the field space because the slow-roll violations in $\eta$ 
become more frequent and large near $N=60$. The isocurvature fluctuations have a negligible growth for $k_1$. The plots for the modes $k_2$ and $k_3$ show that the growth of isocurvature fluctuations start at around $N\sim 70$ for both of them, and the growth of curvature fluctuations follows right away. Inspection of the  background evolution  leads to an intuitive and qualitative understanding of their superhorizon growth at this particular e-fold. At $N= 70.33$,  $\eta$ crosses $-6$ for the first time  and the tangential derivative of the potential, $V_T$, crosses zero for the first time. This corresponds to an ultra slow-roll (USR) regime \cite{Dimopoulos:2017ged,Kinney:2005vj}, as can be seen by using \eqref{varphiT} to write $\eta$  \eqref{etaphi} as follows:
\be
\eta = -6 - \frac{2V_T}{H\dot\varphi} +2\epsilon \,.
\ee
At the same time, $m_T$ starts becoming less tachyonic. After this point during inflation, i.e., for $N \gtrsim 70$, the conditions $\eta \leq -6$ and $V_T=0$ are reached repeatedly during the rest of the course of inflation, owing to the periodic nature of the background quantities. Such an USR condition~\cite{Hooshangi:2022lao} that is attained periodically from $N \sim 70$, may affect the source terms in Eq.s~\eqref{QT} and~\eqref{QN} in such a way that the perturbations start growing while this condition is attained. Inspection of the superhorizon perturbation modes for other parameter combinations presented in the next section also reveals that they all start growing at  a particular e-fold $N_{\rm USR}$ during the course of inflation at and after which the effective USR condition is reached periodically. Both $\eta$ and $V_T$ are background quantities, which determine the starting point of the enhancement of the curvature perturbations for the modes. However, the amount of growth of each perturbation still depends on the turning rates via the Eq.s~\eqref{QT} and~\eqref{QN}. 

The effect of a single sharp turn in the field space for multifield models on the evolution of perturbations have been explored in great detail in the literature \citep{Achucarro:2010da,Achucarro:2010jv,Palma:2020ejf,Fumagalli:2020adf,Chen:2011zf,Shiu:2011qw,Cespedes:2012hu,Achucarro:2012sm,Avgoustidis:2012yc,Gao:2012uq,Achucarro:2012fd,Konieczka:2014zja}. A sharp turn enhances the source terms of the scalar perturbations, and therefore can enhance the power spectrum for modes that are subhorizon during the feature. As a result, inflation models with a sharp turn in field space lead to enhanced adiabatic perturbations, which can be efficient in generating abundant primordial black holes and large amplitude of primordial gravitational waves~\citep{Fumagalli:2021cel,Fumagalli:2020nvq,Ishikawa:2021xya,Ketov:2021fww,Iacconi:2021ltm}. Recently \cite{Boutivas:2022qtl} have explored the effect of controlled number of sharp turns in the field space by modelling the turning rates and obtained enhancement in the curvature power spectra with characteristic oscillatory features that depend on the properties of the feature in turning rates. However, for the case under consideration, the situation is far more complicated due to the presence of multiple features in the background and field space. First of all, the oscillations in the potential make the background parameters deviate from their standard slow-roll evolution ($|\eta |>1$). Secondly, the inflationary trajectory has multiple turns ($\Omega /H$ in Fig.~\ref{bkgd_fig1}) in the field space. Such large number of turns and  intricate features in the field space and its consequences are unavoidable for a viable choice of parameters that can sustain inflation long enough so that the CMB pivot scale $k=0.05$ Mpc$^{-1}$ exits the horizon $\sim 55 - 65$ e-folds before the end of inflation.

An immediate outcome of the above-described features is that the adiabatic power spectrum $P_{\mathcal{R}}(k)$ for this class of models has a non-trivially  enhanced profile. With a judiciously chosen set of parameters, such background oscillations and sharp turns can cumulatively lead to $\sim \mathcal{O}(10^{7})$ enhancement of in the oscillatory envelope of $P_{\mathcal{R}}(k)$ (see Fig.~\ref{fig:pscomp}). The amplitude of $P_{\mathcal{R}}(k)$ depends on $M^2/\beta$, whereas, the peak position and amplitude of the features, as well as details of oscillations in $P_{\mathcal{R}}(k)$ depend on the parameters $\lambda /M$ and $b$. In all of the examples considered in this paper, we keep $\beta =1$ and fix $M$ with the requirement of CMB normalisation at the pivot scale
\footnote{Note that although CMB normalisation is satisfied at the pivot scale $k_{\rm CMB}=0.05$ Mpc$^{-1}$, as mentioned before, it is not consistent with CMB observations for the range of scales probed by CMB due to large oscillations (see Fig.~\ref{fig:pscompcmb} and discussion there). }. 
We note again that the complexity of the cumulative effect of background and field space features makes it difficult to probe the dependence of the growth of $P_{\mathcal{R}}(k)$ on the model parameters. The approach taken here starting from a concrete model of supergravity AM, leads to a featurefull $P_{\mathcal{R}}(k)$ with a peak at scales smaller than those probed at CMB. We note categorically the following:
\bi
\item The enhancement in $P_{\mathcal{R}}(k)$ can be mainly due to multiple sharp turns, however, this is induced by the transient decreases in $\epsilon$ (inflection points) as large as $\epsilon_{\rm max}/\epsilon_{\rm min}\simeq \mathcal{O}(10^{-3})$ (see fig.~\ref{bkgd_fig1}). This is somewhat a combination of the two types of features in the decomposition of $\eta$ presented in~\cite{Boutivas:2022qtl}, where the effect of violations of slow-roll
 in $\delta _{\phi}$ ($-\eta _{\parallel}$  in their notation) and $ \omega$ ($\eta _{\perp}$ in their notation) are shown individually.

\item The small oscillations on top of the enhanced profile are mainly due to the oscillations in the background parameters. However, there can be effects of interference of the characteristic growths due to each of the turns in field space.
\ei
Nevertheless, we have explored a region of the parameter space where (i) inflation is carried out for $55 - 65$ of e-folds; 
(ii) at large scales, the adiabatic power spectra oscillation envelopes have pivot amplitudes $\sim \mathcal{O}(10^{-9})$ and are red-tilted (for some examples);
(iii) $P_{\mathcal{R}}(k)$ is enhanced at small scales with oscillatory features which, for suitable choice of parameters, can lead to interesting outcomes for PBH and GWs. We elaborate these points in detail in the next sections with various values of the control parameters $\lambda /M$ and $b$.

\subsection{Adiabatic power spectrum in multifield axion monodromy}
\label{adpower}

\begin{figure}[H]
\center{
\includegraphics[width=0.49\textwidth]{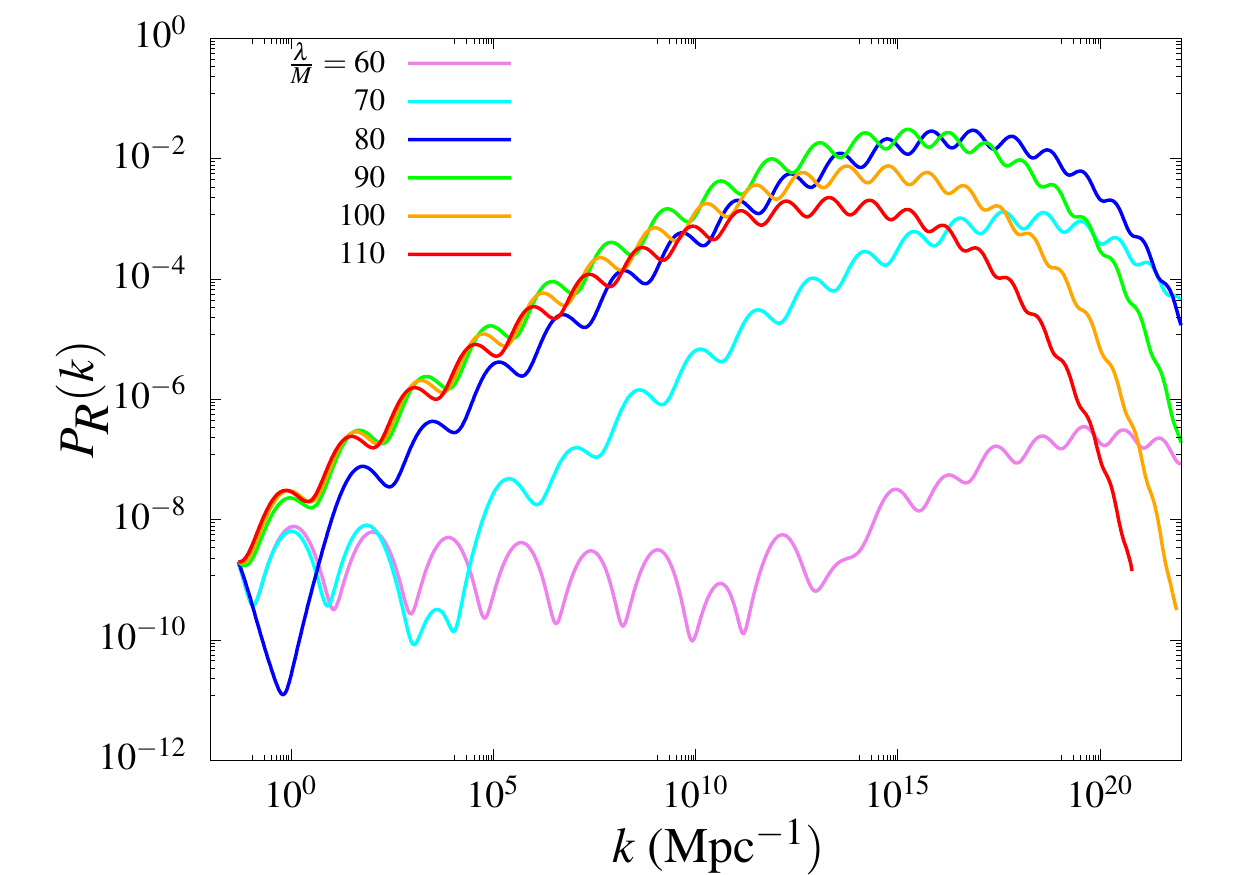}
\includegraphics[width=0.49\textwidth]{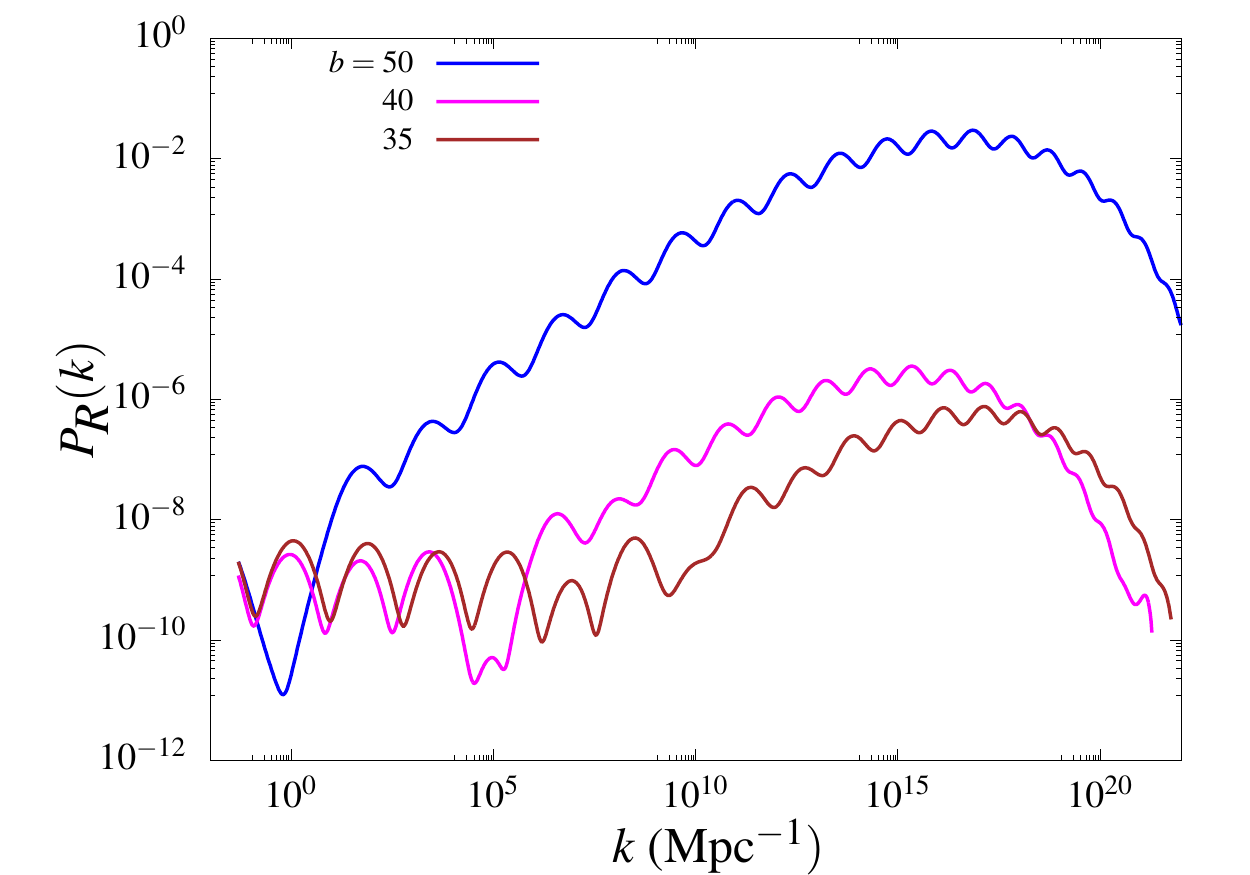}
}
\caption{Adiabatic power spectra for the selection  of parameters given in Table~\ref{tab:2} computed using the code \texttt{PyTransport}. The left panel shows the variation of $P_{\mathcal{R}}(k)$ for different values of $\lambda /M$, with fixed $b=50$. The right panel shows $P_{\mathcal{R}}(k)$ for a fixed $\lambda /M=80$ with varying  $b$. 
In both the panels, the leftmost points of the plots are the CMB pivot scale $k_{\rm CMB}=0.05$ Mpc$^{-1}$.}
\label{fig:pscomp}
\end{figure}

The amplitude of $P_{\mathcal{R}}(k)$ is controlled by an overall parameter $M^2/{\beta}$, whereas the parameters $\lambda /M$ and $b$ determine the oscillatory profile. Interestingly, due to the presence of multiple oscillations in the potential itself for viable parameter combinations, the initial field values also influence slightly the dynamics of inflation. This is due to the fact that for  some  initial values of $\rho$ and $\theta$, one or both of the fields encounter local minima, which makes it difficult to execute slow-roll along that direction.
In Table~\ref{tab:2} we show a suitable set of the parameters and initial conditions   used to  compute  $P_{\mathcal{R}}(k)$ for the supergravity axion monodromy model described above. The perturbation equations~\eqref{QT} and~\eqref{QN} are solved with the transport code \texttt{PyTransport}\footnote{Details about the \texttt{PyTransport} code can be found \href{https://transportmethod.com/pytransport/}{here}.}~\cite{Mulryne:2016mzv} to evaluate $P_{\mathcal{R}}(k)$ for each case shown in Fig.~\ref{fig:pscomp}. 
For a given set of initial values $\rho _i$, $\theta _i$ and the parameters $\lambda /M$ and $b$, the pivot is determined as the point at which the scalar spectral index $n_s$  matches the constraint given by Planck 2018~\cite{Planck:2018jri}: $n_s = 0.9649 \pm 0.0042$ at $68\%$ confidence limit. From the penultimate column of Table~\ref{tab:2}, we see that the tensor-to-scalar ratio $r$ is within the latest bound by BKPlanck 2020~\cite{BICEP:2021xfz}, which is $r< 0.036$ at $95\%$ confidence limit. 
On the other hand,  the oscillations at CMB scales are rather large and thus  violate the Planck bound for an oscillatory power spectrum~\cite{Planck:2018jri,Braglia:2021rej}, as mentioned before. 
These  oscillations can be attributed to the sharp violations of slow-roll and sharp turns in the field space, which seems unavoidable to reach large small-scale power spectra of phenomenological importance.
 
Notice that the values for $r$ do not correspond to either a $\phi^2$-like inflation nor natural inflation as 
the effective decay constants for the examples in Table \ref{tab:2} are of order $f_{\rm eff}\lesssim 10^{-1}M_{\rm Pl}$, and similar to the modulated single field case discussed in  \cite{Ozsoy:2018flq}, the non-perturbative subleading corrections change the background evolution, as well as the cosmological predictions.
 Finally, the parameter $M^2/{\beta}$ can be determined by matching with the pivot amplitude given by Planck 2018. 
\begin{table}[!h]
\centering
\begin{tabular}{|c|c|c|c|c|c|c|c|}
\hline
 \cellcolor[gray]{0.9} $M$ & \cellcolor[gray]{0.9}$\lambda /M$ & \cellcolor[gray]{0.9}$b$ & \cellcolor[gray]{0.9}$\rho _{\rm ini}$ & \cellcolor[gray]{0.9}$\theta _{\rm ini}$&\cellcolor[gray]{0.9}$N_{\rm inf}$ & \cellcolor[gray]{0.9}$r$ & \cellcolor[gray]{0.9}$V_{\rm inf}^{1/4}$ 
\\
\hline
 $2.52 \times 10^{-6}$ & 60 & 50 & 0.250 & 4.20 & 64.77 & 0.010 & 0.0029
\\
\hline
$2.73 \times 10^{-6}$ & 70 & 50 & 0.250 & 4.20 & 62.32 & 0.016 & 0.0030
\\
\hline
$2.15 \times 10^{-6}$ & 80 & 50 & 0.245 & 4.20 & 59.48 & 0.018 & 0.0027
\\
\hline
$6.41 \times 10^{-7}$  & 90 & 50 & 0.250 & 4.20 & 57.49 & 0.020 & 0.0015
\\
\hline
$1.10 \times 10^{-7}$ & 100 & 50 & 0.250 & 4.20 & 56.07 & 0.022 & 0.0006
\\
\hline
$1.25 \times 10^{-8}$ & 110 & 50 & 0.250 & 4.20 & 55.06 & 0.024 & 0.0002
\\
\hline
$1.60 \times 10^{-6}$ & 80 & 40 & 0.250 & 4.50 & 63.63 & 0.011 & 0.0026
\\
\hline
$1.60 \times 10^{-6}$ & 80 & 35 & 0.400 & 5.50 & 56.99 & 0.012 & 0.0026
\\
\hline
\end{tabular}
\caption{Selection   of parameter values in Planck units. We consider $\beta =1$ for all of these sets and fix the CMB normalisation by tuning $M$ only. The number of e-folds from the horizon exit of the pivot scale to the end of inflation is also indicated as $N_{\rm inf}=N_{\rm end}-N_{\rm pivot}$.
}
\label{tab:2}
\end{table}

The parameters $\lambda /M$ and $b$ influence the oscillations in the background dynamics as well as the turns in the field space. Therefore, the position of the peak, $k_p$, and the amplitude at the peak $P_{\mathcal{R}}(k_p)$ also depend on these parameters in a complex manner. The enhancement in $P_{\mathcal{R}}(k)$ at small scales can lead to interesting phenomenological implications which we discuss in the next section. For all the examples in Table~\ref{tab:2}, $c\ll 1$, which leads to a large number of stationary points in the potential for $\rho=\rho_{\rm min}$.

%
The effect of decreasing ``periodicity'' in number of e-folds is clearly inherited in $P_{\mathcal{R}}(k)$, however, the rate of decrease may not follow the same pattern as the background parameters due to the combined contribution of the background effects towards the scalar perturbations.  

It is interesting to note that for the range in $\lambda / M$ considered here, the peak position $k_p$ is maximum for $\lambda / M = 80$. For  $\lambda / M \geq 80$, the dependence of $k_p$ on $\lambda / M$ seems to be mild (left panel of Fig.~\ref{fig:pscomp}), whereas a stronger dependence of $k_p$ on the variation of $b$ can be seen in the right panel of Fig.~\ref{fig:pscomp}.
The mechanism of adiabatic and isocurvature fluctuations sourcing each other is such that the isocurvature power spectra can be large once the growth in curvature perturbations start to set in. However, the isocurvature constraint at CMB scales is checked to be satisfied for each case.

It can be seen from the left panel of Fig.~\ref{fig:pscomp} that $P_{\mathcal{R}}(k)$ has blue-tilt immediately after the pivot scale for $\lambda /M \geq 90$, although constraints on $n_s$ and $r$ are satisfied at the pivot scale. This is related to the mild dependence on initial field values, since once can start evolving a little higher in the plateau, although at the cost of having larger value of $N_{\rm inf}$ and can also violate the CMB constraint on $r$ in some cases. This can have implications for reheating as discussed below.
Finally let us note that the envelope of the modulated power spectrum has an approximate $k^2$ behaviour for all the cases considered. Also for the tiny oscillations in the power spectra, the average slope is $k^2$, whereas it can reach $\sim k^3$ very few times in some of the cases.
\subsection*{Power spectrum at CMB scales}
\begin{figure}[H]
\center{
\includegraphics[width=0.49\textwidth]{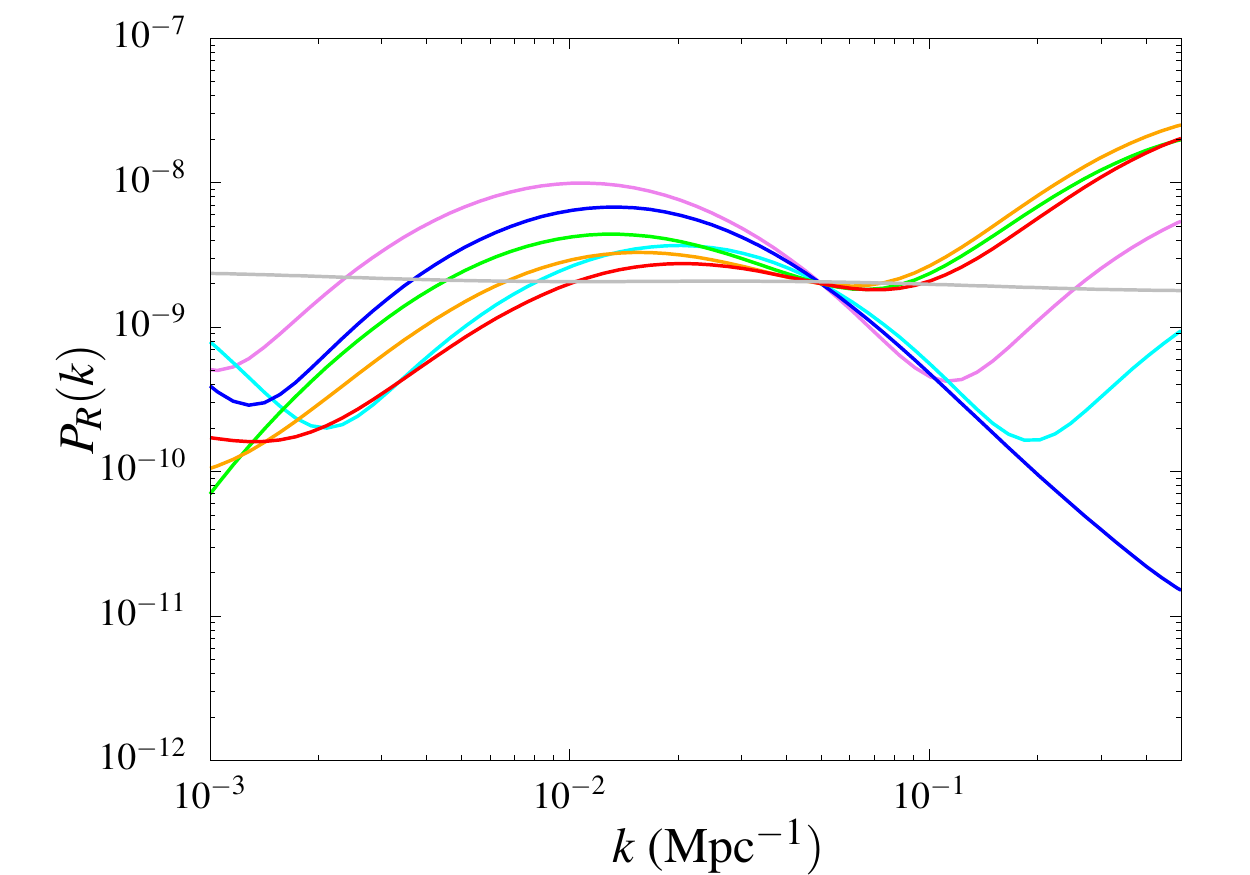}
\includegraphics[width=0.49\textwidth]{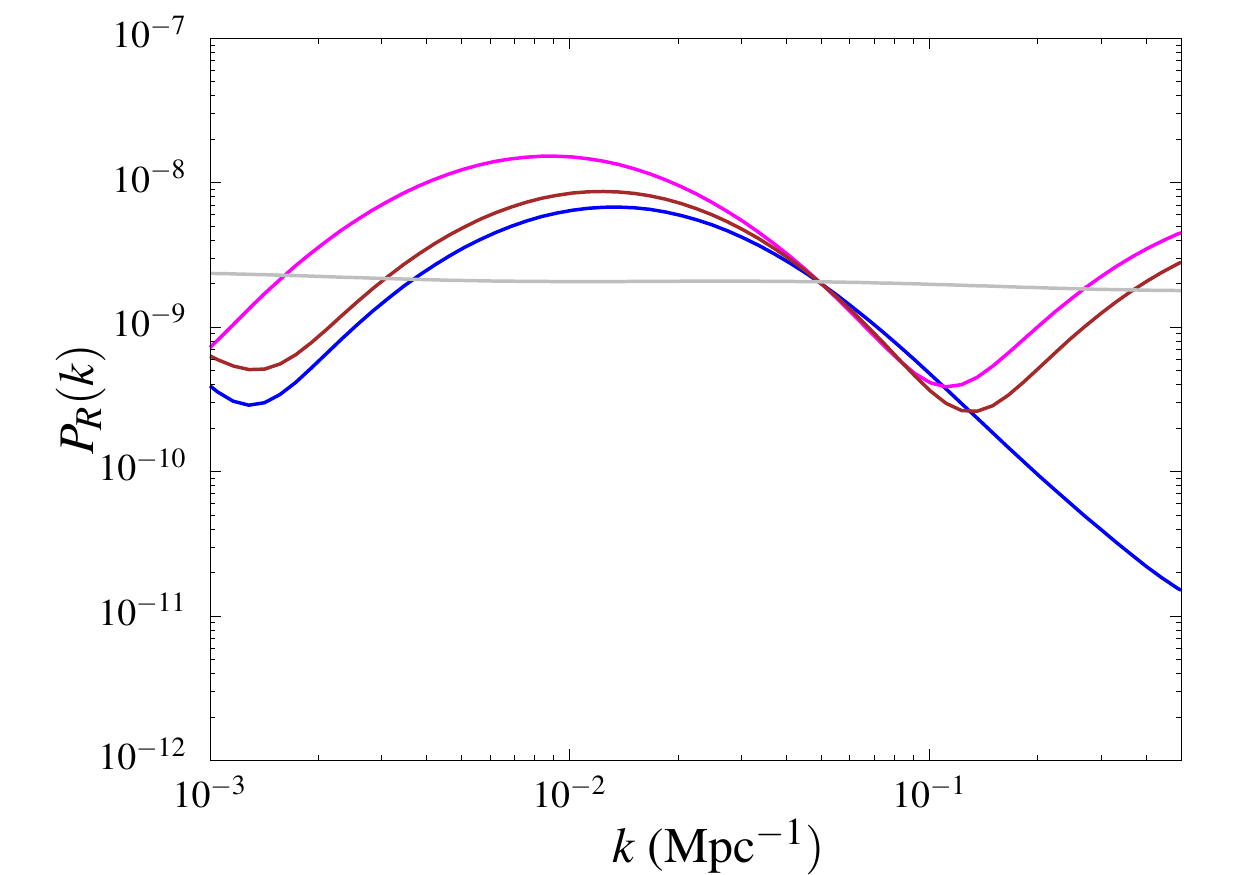}
}
\caption{Adiabatic power spectra around the CMB scales for the selection  of parameters given in Table~\ref{tab:2} computed using the code \texttt{PyTransport}. The left panel shows the variation of $P_{\mathcal{R}}(k)$ for different values of $\lambda /M$, with fixed $b=50$. The right panel shows $P_{\mathcal{R}}(k)$ for a fixed $\lambda /M=80$ with varying  $b$. The color code is the same as in Fig.~\ref{fig:pscomp}. }
\label{fig:pscompcmb}
\end{figure}

The large oscillations in $P_{\mathcal{R}}(k)$ result  from the oscillations present in the background parameters, as well as  to the sharp turns. For the set of parameters considered in Table~\ref{tab:2}, these oscillations are very large even at the CMB scales, as can be seen in Fig.~\ref{fig:pscompcmb}. 
Although, at the CMB pivot scale $k_{\rm CMB}=0.05$ Mpc$^{-1}$, and only at that scale, the quantities $n_s$ and $r$ are within the $1\sigma$ confidence level, 
inspecting the highly oscillating power spectra around the pivot scale reveals that the model with the  set of fiducial values for the model parameters in Table~\ref{tab:2}, is outside the observational CMB bounds on these scales (indicated by the grey line in the figure). 
Planck 2018~\cite{Planck:2018jri} constrains the oscillatory power spectra from inflation by limiting the fractional amplitude of oscillations. For a power spectrum with logarithmic oscillations of the form:
\begin{equation}
P_{\mathcal{R}}(k)=P_0 (k) [1+A_{\rm log}\cos (\omega _{\rm log}\log \kappa + \phi _{\rm log})],
\label{Pkplancklog}
\end{equation}
where $\kappa = k/k_{\rm CMB}$.
Planck 2018 presents a marginalised confidence levels in the $A_{\rm log} - \omega _{\rm log}$ space, where the limit on the fractional amplitude for oscillation is $A_{\rm log}\lesssim 0.03$ (see Sec. 7.1.1 and Fig.(28) in reference~\cite{Planck:2018jri}).
For the multifield AM inflation model in the present work, $P_{\mathcal{R}}(k)$ has logarithmic oscillations in $k$, although the exact fitting is more complicated than the form in Eq.~\eqref{Pkplancklog}. However,
from Fig.~\ref{fig:pscompcmb}, it is evident that the oscillations are too large to be consistent with the Planck value for the fractional amplitude of oscillation. 

 For a model with resonant features at small scales, CMB consistency could be achieved  for  oscillations growing with time, such that they are small enough at CMB scales, to be consistent with the constrained value of $A_{\rm log}$, and larger at small scales. 
 For our present model, 
 the fractional oscillation amplitude remains large and nearly constant for the whole range of scales, which can be seen from Figs.~\ref{fig:pscomp} and~\ref{fig:pscompcmb}. Choosing parameters  such as to make $P_{\mathcal{R}}(k)$ CMB consistent makes the  growth feature at small scales too small, which makes the resonant oscillations uninteresting in the context of PBH and induced GW. 
This indicates how one can modify the scalar potential to ensure these requirements are achieved and we leave such a study for future work.

\subsubsection*{Implications for reheating}

We see from Table~\ref{tab:2}  that in order to account for a large enough amplification in the scalar power spectrum at small scales 
the observable scales associated with the CMB have to leave the horizon $N_{\rm inf}=N_{\rm end} - N_{\rm pivot}$ e-folds before the end of inflation. These values have non-trivial theoretical implications on the reheating phase after inflation.

In general, there is a theoretical uncertainty in determining $N_{\rm inf}$ due to the unknown thermal history of the Universe after inflation. This ignorance is usually parameterised by an effective equation of state (e.o.s.) $w_{\rm re}$ and the number of e-foldings during the reheating epoch $N_{\rm re}$ using the matching equation \cite{Adshead:2010mc,Maharana:2017fui}
\be\label{Ninf1}
N_{\rm inf} \simeq 57+\frac{1}{4}\ln r-\frac{1}{4}(1-3w_{\rm re})N_{\rm re}. 
\ee
Usually,  $0\lesssim w_{\rm re}<1/3$, although other exotic scenarios may arise, such as $w_{\rm re}=1$. However if a non-standard cosmological history prior to BBN arises, for example due to a scalar dominated epoch, it induces a further modification to  \eqref{Ninf1}  \cite{Maharana:2017fui}, 
$\delta N_{\rm inf} \equiv \frac{1}{4}\ln \Gamma$, where $\Gamma$ parameterises the change to $N_{\rm inf}$ due to this non-standard epoch. In was shown in  \cite{Maharana:2017fui} that this modification can span a large range of positive values, depending on the scalar model and the reheating temperature. Thus we see that accounting for the $N_{\rm inf}$ in some of the examples  in Table \ref{tab:2} in case we proceed to obtain red-tilt of the power spectra at large scales (e.g.~$\lambda/M=90, b=50$) imply a relatively long period of scalar field domination.

\section{Primordial black holes and primordial gravitational waves}
\label{PBHPGW}
When primordial adiabatic fluctuations of large amplitude re-enter the horizon in the post-inflationary era, they can collapse gravitationally to form black holes (BH). These primordial black holes (PBHs) can form a part or all of the observed dark matter abundance depending on the PBH mass range. If PBHs exist, they can affect several astrophysical and cosmological phenomena, such as lensing, galactic and extra-galactic radiation etc.~and therefore, their observation can put constraints on the abundance of PBHs as dark matter over a large range of PBH masses. In particular, observations of massive BH binary mergers in LIGO/Virgo surveys have rekindled the enthusiasm about models of inflation that can lead to large scalar fluctuations at small scales and therefore abundant PBHs. 

A variety of observational and experimental data now constrain a significant part of the PBH parameter space. Nevertheless, these constraints are expected to evolve in the near future with the prospect of additional data and improved analysis. PBHs evaporate on a timescale $t_{\rm ev}=5120\pi G^2M^3/(\hslash c^4)$ via Hawking radiation, and therefore PBHs of mass lower than $M\simeq 5\times 10^{14}~{\rm g}\simeq 2.5\times 10^{-19}\ms$ have completely evaporated by now\,\cite{Hawking:1974rv}. Slightly heavier PBHs have not completely evaporated yet and may radiate gamma-ray photons, neutrinos, gravitons and other massive particles at different stages of evaporation. Therefore, constraining the injection of photons and neutrinos in the (extra-)galactic medium using Voyager data, extra-galactic radiation background, SPI/INTEGRAL observations, etc.~\cite{Churazov:2010wy,Siegert:2016ijv,Laha:2019ssq,Bays:2011si,Collaboration:2011jza,Agostini:2019yuq,Dasgupta:2019cae,Laha:2020ivk}, the abundance of light PBHs with $M \lesssim 10^{-17}\ms$ can be constrained. CMB anisotropies and abundance of light elements at the time of nucleosynthesis due to the energy decomposition in the background by the evaporation products from the black holes~\cite{Acharya:2020jbv} can constrain PBHs for  masses $M \geq 5.5\times 10^{-21} \ms$ and  $M \simeq 10^{-22} - 10^{-21}\ms$ respectively. PBHs in the mass range $10^{-11}\ms<M<10^{-1}\ms$ are constrained by their gravitational lensing of light rays from distant stars. Observation of the stars in the M31 galaxy by the HSC telescope, the EROS and OGLE survey together now rule out the contribution of PBH towards total DM density above $1 - 10\%$ in this mass range\,\cite{Smyth:2019whb,Tisserand:2006zx,Niikura:2019kqi}. The caustic crossing event for the star Icarus or MACS J1149LS1 and the resultant strong lensing has been used to place constraint on compact objects in the range $10^{-5}\ms<M\lesssim 10^3\ms$\,\cite{Oguri:2017ock}. The GW detections by the LIGO/Virgo collaboration put an upper bound on $\fpbh$ in the mass region $0.2\ms<M<300\ms$ assuming that the observed binary BH mergers are PBH mergers in early or late Universe\,\cite{Ali-Haimoud:2017rtz,Bird:2016dcv,Sasaki:2016jop,Cholis:2016kqi,Clesse:2016vqa,Raccanelli:2016cud,Kovetz:2017rvv,Authors:2019qbw,Kavanagh:2018ggo,DeLuca:2020qqa,Wang:2016ana}. Finally, the radiation from the accreted gas around PBHs of mass $M \gtrsim 100\ms$ affects the spectrum and the anisotropies of the CMB\,\cite{carr1981pregalactic,Ricotti:2007au,Serpico:2020ehh}.

Evidently, current observations have already constrained the PBH population in a significant part of its mass range meaning that it cannot form more than a few percent of total DM abundance in a large part of the mass range. However, most of these constraints depend on the width of the PBH mass spectrum and are usually quoted for a monochromatic mass spectrum ($\Delta M \sim M$)\footnote{For detailed reviews on the current constraints on PBH abundance, see Refs.~\cite{Carr:2020gox,Green:2020jor,Carr:2021bzv}. 
See also \cite{Khlopov:2008qy} for an earlier review.}.
 
If inflation generates scalar fluctuations large enough to produce abundant PBH, then they can also lead to large amplitude of secondary gravitational waves (GW) owing to the coupling between the scalar and tensor modes in second and higher orders of perturbation theory~\cite{Baumann:2007zm,Espinosa:2018eve,Kohri:2018awv,Ananda:2006af,Domenech:2021ztg}. These scalar-sourced GWs appear as a stochastic background today (SGWB). Since the scalar source depends on the scalar power spectrum as $\mathcal{S}_\textbf{k} \propto P_{\mathcal{R}}(k)^2$ (see Sec.~\ref{secGW}), for a featureless red-tilted $P_{\mathcal{R}}(k)$ the amplitude of SGWB is tiny. However, if $P_{\mathcal{R}}(k)$ is enhanced sufficiently, which is generally the case for inflation models with small-scale features, then the resulting GW spectrum $\Omega _{\rm GW}$ can be large enough to be possibly detected in the ongoing and upcoming GW surveys. The frequency $f$ of this induced GW depends on the mode $k$ entering the horizon at the post-inflationary time when the GW is sourced. In the radiation dominated (RD) epoch, 
\begin{equation}
f=\frac{kc}{2\pi a_0}=1.5\times 10^{-15} k {~\rm Mpc} {~\rm Hz}.
\label{fk}
\end{equation} 
The present and proposed GW surveys span over decades in the frequency space. Pulsar timing arrays (PTAs) such as NANOGrav~\cite{Aggarwal:2018mgp,NANOGrav:2020bcs}, EPTA~\cite{Lentati:2015qwp,Shannon:2015ect,Qin:2018yhy} etc., are sensitive in the range $10^{-9} - 10^{-7}$ Hz, corresponding to $6\times 10^5 {~\rm Mpc}^{-1}\lesssim k \lesssim 6\times 10^7 {~\rm Mpc}^{-1}$. Ground based interferometric detectors such as LIGO/Virgo~\cite{LIGOScientific:2019vic,LIGOScientific:2016aoc,LIGOScientific:2016dsl,LIGOScientific:2017ycc}, KAGRA~\cite{Akutsu:2015hua,Haino:2020age} and ET~\cite{Maggiore:2019uih} cover the range $10 - 10^{3}$ Hz, corresponding to $6\times 10^{15} {~\rm Mpc}^{-1}\lesssim k \lesssim 6\times 10^{18} {~\rm Mpc}^{-1}$. The intermediate frequency range can be probed by LISA~\cite{LISA:2017pwj,Kaiser:2020tlg,Barausse:2020rsu,LISACosmologyWorkingGroup:2022kbp}, DECIGO~\cite{Seto:2001qf,Yagi:2011wg,Kawamura:2020pcg}, AION/MAGIS~\cite{Badurina:2019hst}, Taiji~\cite{Ruan:2018tsw}, TianQin~\cite{TianQin:2015yph}.

The supergravity axion monodromy model studied in this work proves to be an interesting candidate to produce abundant PBHs and large secondary GWs with a characteristic profile as we see below. It can be seen in the $P_{\mathcal{R}}(k)$ curves in Fig.~\ref{fig:pscomp} that $P_{\mathcal{R}}(k)$ can reach large amplitudes (up to $10^{-3} - 10^{-2}$ in some cases)  at small scales for multiple parameter sets considered in Table \ref{tab:2}. 

\subsection{PBH formation}
\label{secPBH}
If the amplitude of the primordial fluctuations is such that when the modes re-enter the horizon at the post-inflationary epoch, the density fluctuations $\delta$ are larger than the critical density for collapse ($\delta_c$), then PBHs can be produced with mass $M_{\rm PBH}=\gamma M_H$, where $M_H$ is the horizon mass at collapse. $\gamma$ signifies the efficiency of collapse~\cite{Carr:1975qj} and is typically of order $1$; here we consider $\gamma = 0.33$. The mass of PBH depends on the horizon size at the time of collapse. For standard post-inflationary history, PBHs with masses of interest are formed in the radiation dominated (RD) epoch ($\delta _c = 0.41$) when the time of collapse can be considered the same as the time of horizon entry for the scalar mode due to logarithmic growth of subhorizon perturbations. 
Using the Press-Schechter formalism for gravitational collapse, PBH abundance at present in an interval of mass $M_{\rm PBH}$ to $M_{\rm PBH}+dM_{\rm PBH}$ produced in RD epoch is 
\begin{equation}
\psi(M_{\rm PBH})=\frac{\gamma}{T_{\rm eq}}\bigg(\frac{g_s(T_1)}{g_s(T_{\rm eq})}\bigg)^{\frac{1}{3}}\bigg(\frac{\Omega_m h^2}{\Omega_c h^2}\bigg)\bigg(\frac{90\Mp^2}{\pi ^2g_*(T_1)}\bigg)^{\frac{1}{4}}(4\pi \gamma \Mp^2)^{\frac{1}{2}}\frac{\beta_{\rm PBH} (M_{\rm PBH})}{M_{\rm PBH}^{\frac{3}{2}}}.\label{psiM}
\end{equation}
The PBH mass fraction is defined as the fraction of the energy that collapses to a PBH at the time of formation, which, for Gaussian adiabatic fluctuations, can be written as
\begin{equation}
\beta_{\rm PBH}(M_{\rm PBH})={\rm erfc} \left[\frac{\delta_c}{\sqrt{2\sigma_{\delta}^2}}\right], \label{beta_w}
\end{equation}
where $\sigma_{\delta}^2$ is the variance of the density power spectrum and calculated as: 
\begin{equation}
\sigma_{\delta}^2=\frac{16}{81}\int \frac{dk}{k}(kR)^4 W^2(k,R)P_{\mathcal{R}}(k).
\label{vardel}
\end{equation}
The window function $W^2(k,R)=\exp(-k^2R^2)$ is chosen to smooth the perturbations on the comoving scale $R$ at formation. The mass $M_{\rm PBH}$ of the PBH produced is related to the comoving wavenumber $k$ via 
\begin{equation}
M_{\rm PBH}(k)=4\pi\gamma {\Mp}^2\bigg(\frac{\pi ^2 g_*^{\rm eq}}{45\Mp^2}\bigg)^{\frac{1}{2}}(a_{\rm eq}T_{\rm eq})^2k^{-2}. \label{Mk_w} 
\end{equation}
$\psi(M_{\rm PBH})$ can be calculated from Eq.~\eqref{psiM} using Eq.s~\eqref{beta_w},~\eqref{vardel} and~\eqref{Mk_w}. The fraction of DM as PBHs for a wide mass spectrum $\psi(M_{\rm PBH})$ can then be defined as 
\begin{equation}
f_{\rm PBH} \equiv \frac{\Omega _{\rm PBH}}{\Omega _{\rm DM}} = \int \psi (M_{\rm PBH}) dM_{\rm PBH},
\end{equation}
which is also dubbed as the PBH abundance.
\begin{figure}[h]
\center{
\includegraphics[width=0.6\textwidth]{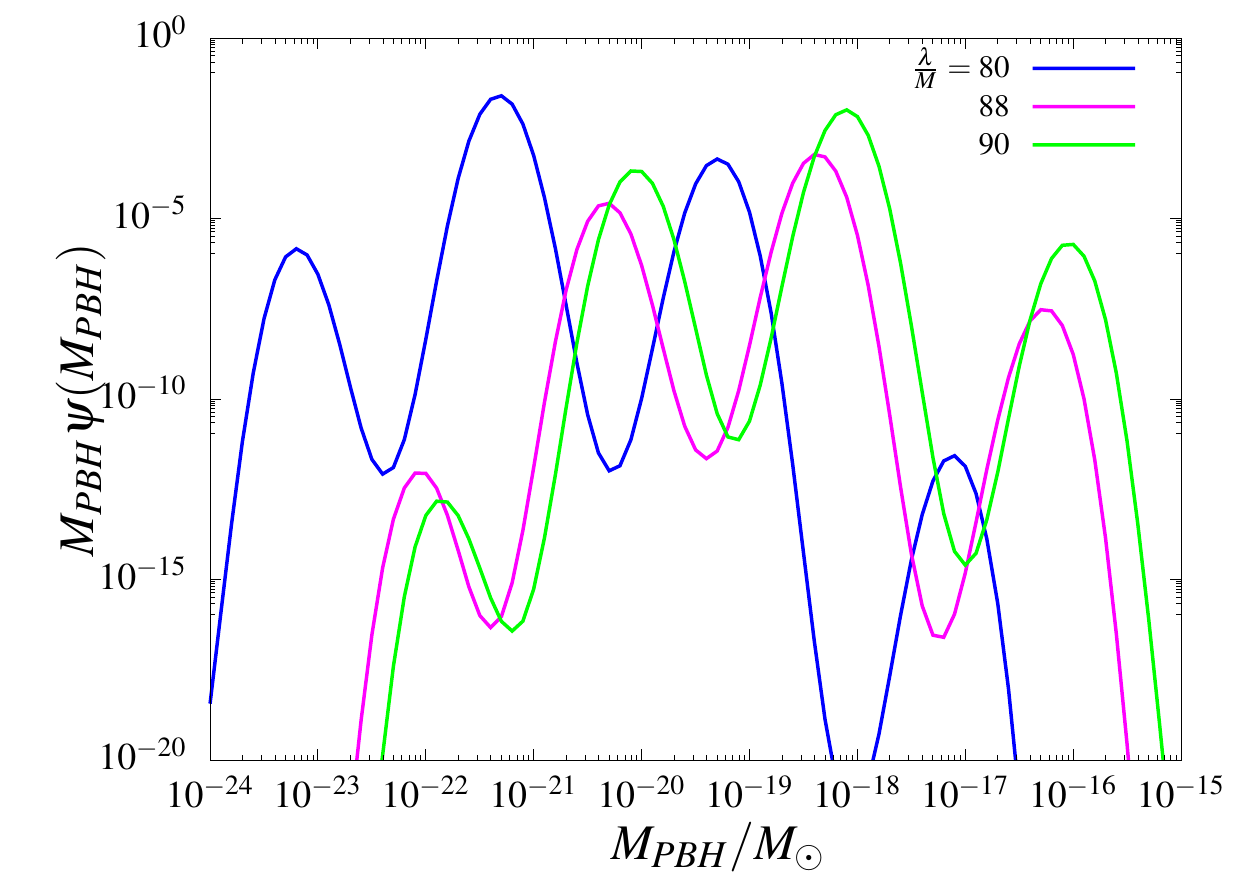}
}
\caption{PBH mass spectra for different choices of $\lambda/ M$ with $b=50$ in all the cases. }%
\label{fig:PBHcomp}
\end{figure}
It can be shown that, typically, to have a considerable PBH abundance in RD, $P_{\mathcal{R}}(k)$ is needed to be enhanced by seven orders of magnitude in amplitude as compared to its CMB value. From the $P_{\mathcal{R}}(k)$ curves plotted in Fig.~\ref{fig:pscomp}, evidently, this criteria is fulfilled for a reasonable  parameter space of the inflation model. 

From the left panel of Fig.~\ref{fig:pscomp}, it can be seen that for $\lambda /M \sim 80 - 90$, the peak amplitude of $P_{\mathcal{R}}(k)$ reaches $\gtrsim 10^{-2}$. With this motivation, in Fig.~\ref{fig:PBHcomp} we present the PBH mass spectra for $\lambda /M \sim 80$, $90$ and $88$ with $b=50$ for all the cases. The first two parameter sets are detailed in Table~\ref{tab:2}, whereas for the last set, similarly, we take $\beta =1$ and $M$ is determined to obey the CMB normalisation. The mass spectra in Fig.~\ref{fig:PBHcomp} exhibit multiple peaks, which can be attributed to the oscillations in $P_{\mathcal{R}}(k)$. 

All of the mass spectra in Fig.~\ref{fig:PBHcomp} peak at very small masses in the range $10^{-24}\ms - 10^{-16}\ms$, which has strong bounds from evaporation constraints from BBN~\cite{Carr:2009jm}, CMB spectral distortions and anisotropies~\cite{Acharya:2020jbv,Chluba:2020oip}, extragalactic $\gamma-$rays~\cite{Carr:2009jm}, Galactic $\gamma-$rays \cite{Carr:2016hva} and Voyager-1 $e\pm$~\cite{Boudaud:2018hqb} (see Figs 4 and 11 in~\cite{Carr:2020gox}). 
These light PBHs can, in turn, induce GWs via poisson fluctuations~\cite{Papanikolaou:2020qtd,Papanikolaou:2022chm} or via Hawking evaporation~\cite{Ireland:2023avg}.
These multiple peaks in PBH mass spectra are interesting, since they can lead to abundance of PBHs in specific narrow mass ranges, while still keeping the total abundance $\fpbh$ small. We found $\fpbh =1.8\times 10^{-2}$ for $\lambda /M =80, b=50$;  $\fpbh = 5.0\times 10^{-4}$ for $\lambda /M =88, b=50$ and $\fpbh = 2.5\times 10^{-3}$ for $\lambda /M =90, b=50$. While BBN constraint still allows $\fpbh \sim 10^{-4}$, CMB and $\gamma-$ray observations constrain $\fpbh$ very stringently below $10^{-10}$ for the range $10^{-20}\ms \lesssim M_{\rm PBH} \lesssim 10^{-17}\ms$ for a monochromatic $\psi (M_{\rm PBH})$. 

Since the oscillations in $P_{\mathcal{R}}(k)$, and therefore the positions and heights of the peaks in the mass spectra are controlled by $\lambda /M$ and $b$, a suitable combination of these parameters can lead to a $M_{\rm PBH}\psi (M_{\rm PBH})$ profile and $f_{\rm PBH}$ that satisfy the observational constraints. It is also worth mentioning here that PBH mass spectra with multiple narrow peaks mean that PBHs of specific masses are produced far more abundantly than others. Given the non-uniform nature of the observational constraints on PBH over the mass range, such specific production of PBHs can lead to interesting phenomenologies, once their exact dependence on model parameters is well understood.

\subsection{Secondary GW}
\label{secGW}
In the second order of perturbation theory, the tensor modes are sourced by first order adiabatic perturbations via the following equation for the tensor mode $k$
\begin{equation}
\label{equ:hhh}
h_k'' + 2 {\cal H} h_k' + k^2 h_k=  4{\cal S_{\mathbf{k}}}(\tau)\, , 
\end{equation}
where ${\cal H}=aH$. The source term ${\cal S}(\mathbf{k}, \tau)$ is calculated at the conformal time $\tau$ in RD epoch\footnote{All the expressions in this text are written explicitly for RD epoch. For secondary GW production during a general epoch~\cite{Allahverdi:2020bys}, see \cite{Domenech:2019quo,Kohri:2018awv,Domenech:2021ztg,Bhattacharya:2019bvk} and the references therein.} can be written in terms of the gravitational potential $\Phi _{\textbf{q}}$ as
\begin{equation}
{\cal S}(\mathbf{k}, \tau)=\int \frac{d^3q}{(2\pi)^{3/2}}e_{ij}(\textbf{k}) q_iq_j\bigg( 2\Phi _{\textbf{q}}\Phi _{\textbf{k-q}}+(\mathcal{H}^{-1}\Phi '_{\textbf{q}}+\Phi _{\textbf{q}})(\mathcal{H}^{-1}\Phi '_{\textbf{k-q}}+\Phi _{\textbf{k-q}})\bigg).\label{source}
\end{equation}
$\Phi _{\textbf{q}}$ are the scalar modes at the time of horizon entry of the momentum $\textbf{q}$ so that $\Phi _{\textbf{q}}=\Phi (q\tau)\frac{2}{3}\mathcal{R} _{\textbf{q}}$, where $\mathcal{R}_{\textbf{q}}$ are the primordial scalar perturbations and $\Phi (q\tau)$ is the scalar transfer function which follows the following evolution in a RD epoch
\begin{equation}
\Phi ''(q\tau) + \frac{4}{\tau}\Phi '(q\tau)+\frac{q^2}{3}\Phi (q\tau)=0. \label{scalaevol}
\end{equation}
Therefore, the source function can be obtained from the scalar power spectrum since
\begin{equation}
\mathcal{P}_{\mathcal{R}}(k)\equiv \frac{k^3}{2\pi ^2}\delta (\textbf{k} + \textbf{k}') \langle \mathcal{R} _{\textbf{k}}\mathcal{R} _{\textbf{k}'}\rangle.
\end{equation}
The scalar induced second order tensor power spectrum is~\cite{Kohri:2018awv},
\begin{align}\label{eq:pgamma}
\overline{\mathcal{P}_h(k,\tau)}=4\int_0^\infty dv \int_{|1-v|}^{1+v} du\left[\frac{4v^2-\left(1+v^2-u^2\right)^2}{4uv}\right]^2 P_{ \mathcal{R}}(kv)P_{\mathcal{R}}(ku)\overline{I^2}(v,u,x)\,,
\end{align}
where the bar denotes the oscillation average.
The momenta $\textbf{q}$ and $\textbf{k}$ are reparameterised in terms of convenient variables $v\equiv q/k$, $u\equiv|\mathbf{k}-\mathbf{q}|/k$, and $x=k\tau$. The integration kernel $I(u,v,x)$ contains the source information (for details, see~\cite{Espinosa:2018eve,Kohri:2018awv}) which can be written with a simple analytical expression for late time, i.e. $x \gg 1$. For horizon entry in the RD era,
\begin{align}
\overline{I_{RD}^2}(x\gg 1,u,v)=\frac{9(u^2+v^2-3)^2}{16u^6v^6x^2}&\Bigg\{ {\pi}^2(u^2+v^2-3)^2\Theta(u+v-\sqrt{3}) \nonumber\\&+  \bigg((-4uv+(u^2+v^2-3) \ln {\frac{3-(u+v)^2}{3-(u-v)^2}}^2 \bigg)\Bigg\}\,.
\label{IRD}
\end{align}
\begin{figure}[H]
\center{
\includegraphics[width=0.48\textwidth]{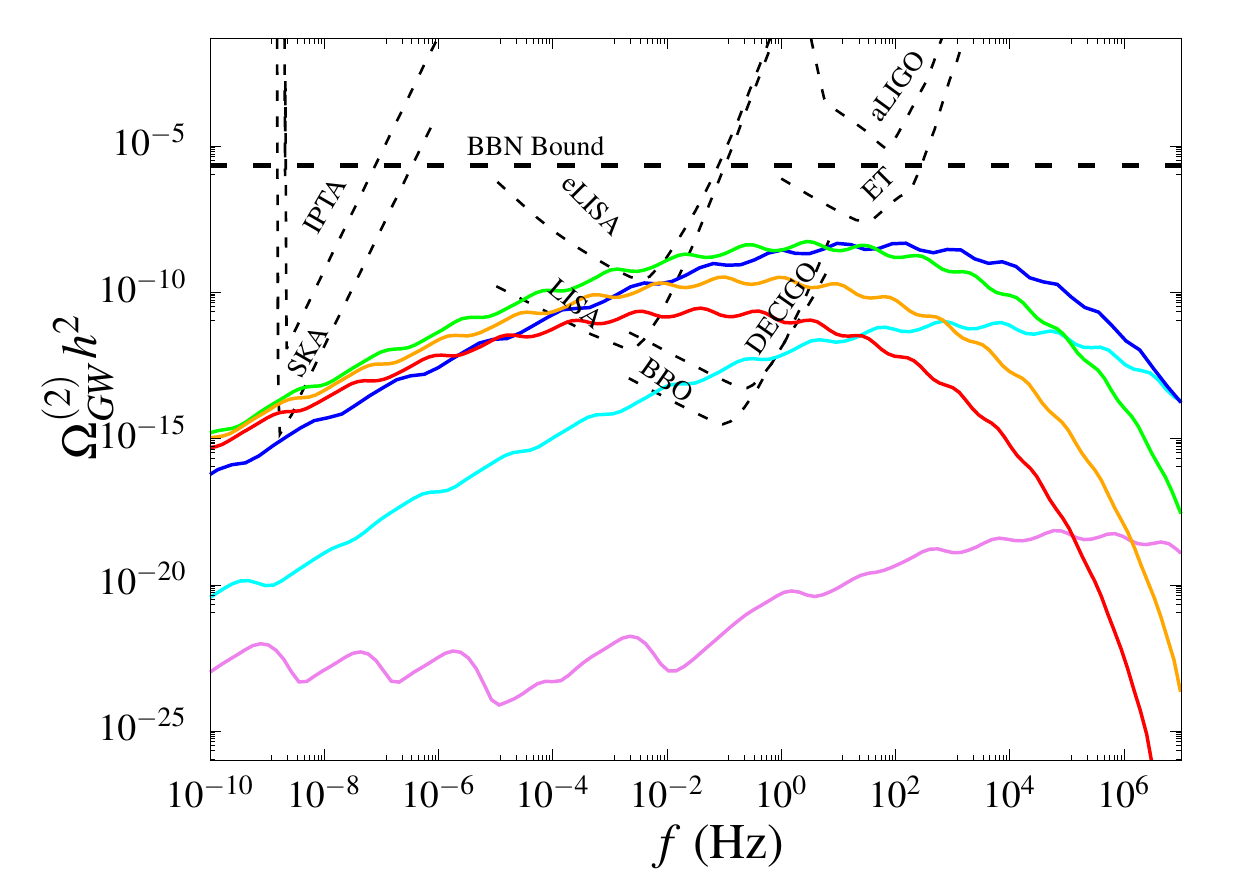}
\includegraphics[width=0.48\textwidth]{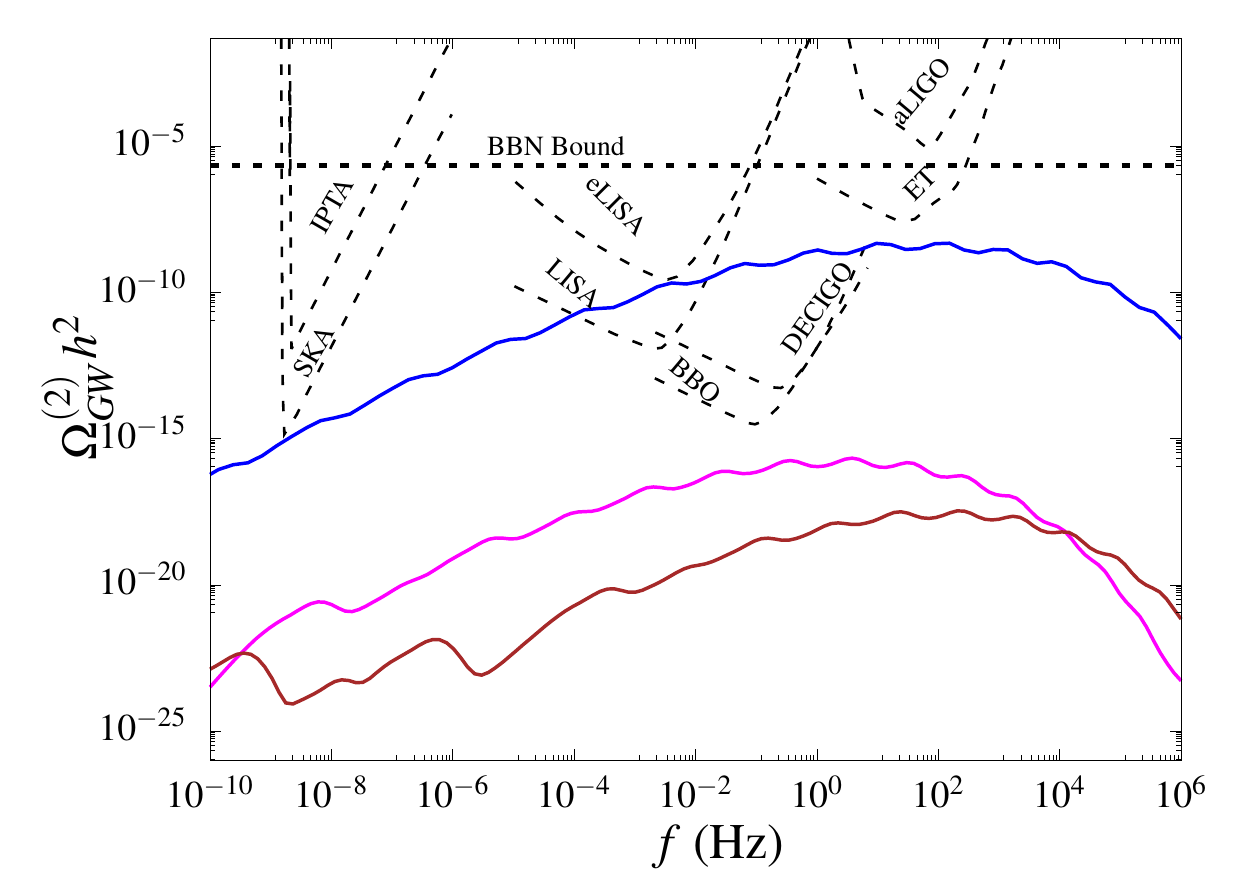}
}
\caption{GW spectra for different choices of parameters given in Table~\ref{tab:2}, where the color schemes are same as in Fig.~\ref{fig:pscomp}. The left panel shows the variation of $\Omega _{\rm GW}^{(2)}h^2$ for different values of $\lambda /M$, with fixed $b=50$. The right panel shows $\Omega _{\rm GW}^{(2)}h^2$ for a fixed $\lambda /M=80$ with varying $b$. }
\label{fig:GWcomp}
\end{figure}
The GW spectrum at present time $\tau _0$ for this secondary GW background is then
\begin{equation}
 \Omega_{{\rm GW}}^{(2)}(k,\tau _0)= 1.62\times 10^{-5}\frac{\Omega _{{\rm rad},0}}{4.18\times 10^{-5}} \bigg(\frac{g_*(\tau)}{106.75}\bigg)\bigg(\frac{g_s(\tau)}{106.75}\bigg)^{-4/3} \Omega_{{\rm GW}}^{(2)} (k,\tau ),
\label{omg2fullnow}
\end{equation}
where 
\begin{equation}
 \Omega_{{\rm GW}}^{(2)} (k,\tau )=\frac{(k\tau )^2}{24} \overline{\mathcal{P}_h(k,\tau)}. \label{omg2trans}
\end{equation}
Due to the form of the kernel given in Eq.~\eqref{IRD}, the induced tensor power spectrum $\overline{\mathcal{P}_h(k,\tau)}$ gathers a power $(k\tau )^{-2}$, therefore, $\Omega_{{\rm GW}}^{(2)}(k,\tau _0)$ is independent of $\tau$. 
Hence, the full second order GW can be calculated using Eq.~\eqref{omg2fullnow} once the primordial power spectrum $P_{\mathcal{R}}(k)$ is obtained for the model under consideration. Fig.~\ref{fig:GWcomp} shows the induced secondary GW spectra the  parameter sets in Table~\ref{tab:2} for the supergravity axion monodromy model considered in this paper.

As expected from the wide enhancement profiles in $P_{\mathcal{R}}(k)$, the GW spectra have wide peak profiles with inherent oscillations. Interestingly, for some of the parameter combinations in Table~\ref{tab:2}, $\Omega _{\rm GW}^{(2)}h^2$ crosses the sensitivity bounds of more than one GW survey at various frequencies. For example, the green, orange and red curves in the left panel of Fig.~\ref{fig:GWcomp} with $\lambda / M \geq 90$ are within the sensitivities of SKA, LISA and DECIGO. The blue curve with $\lambda / M = 80$ is within the sensitivity of LISA and DECIGO. The possibility of simultaneous detection at different observations is encouraging, since cross-correlation of between these surveys can put stringent constraints on the model parameters in such cases, even for non-detection of such GW profiles.
Moreover, the non-trivial spectral shape and amplitude of a the GW of this class of models, may be detectable by LISA~\cite{Caprini:2019pxz}. Finally, let us note that despite the wide profile of the GW spectra, the BBN bound $\int \frac{df}{f} \Omega _{\rm GW}h^2 < 5.6\times 10^{-6}\Delta N_{\rm eff}$ is satisfied for all of the examples considered here. 

\section{Discussion and Conclusions}\label{Sec4}

In this work, we presented a novel mechanism to generate transient turns in the field space of multifield inflation using transient violations of slow-roll for multifield axion monodromy inflation in supergravity, without the need of large curvatures  of the field space. 
Modulations in the scalar potential is a generic feature of this model due to non-perturbative terms. The saxion and axion are coupled non-trivially, which then gives rise to a non-trivial inflationary evolution and attractive cosmological implications.  
The effect due to the modulations in the background evolution is shown in figures \ref{fig:fieldomega} and \ref{bkgd_fig1}, which show that transient slow-roll violations induce transient  large turning rates, even with order one scalar field curvature (${\mathbb R}=-4$), as typically arise in supergravity and string theory. 

The final effect on the scalar perturbations is interesting with an enhanced profile for small scales with resonant oscillations. 
The effect of a single turn in the field space has been studied extensively and analytically in literature, albeit with the turn introduced ad hoc. An analytical study for the detailed effects of multiple (5-6) turns in the field space has been presented in~\cite{Boutivas:2022qtl}.
Here, we take a different approach to incorporate turns in field space via the slow-roll violations as a virtue of the potential itself. Indeed, a detailed (possibly analytical) study of the dependence of the background and perturbations on the model parameters might be possible if one could  single out one turn as the major contributor. Nonetheless, in our multifield axion monodromy model, multiple turns with comparable turning rates are unavoidable. Thus,  we calculated numerically  the adiabatic power spectra using the \texttt{PyTransport} code.

We have shown in section~\ref{adpower} that $P_{\mathcal{R}}(k)$ inherits the background oscillations and it can have an enhanced peak profile at small scales. Although the exact dependence of the $P_{\mathcal{R}}(k)$ profile on the slow-roll parameters (and by extension, the potential parameters) 
is difficult to estimate analytically,  the profile can be attributed to the shape and duration of the turns in the field space as well as their presence in large numbers. The extrinsic dependence of  $P_{\mathcal{R}}(k)$ on the parameters $\frac{\lambda}{M}$ and $b$ can be seen in Fig.~\ref{fig:pscomp}.
It is to be emphasised that even though we quote the initial values for the fields $\rho$ and $\theta$, this model is not fine-tuned, as long as a reasonable number of inflationary e-folds is ensured. 
However, as we discussed, the oscillations are large around CMB scales and the fractional amplitude of oscillations is thus in tension with Planck bounds for oscillatory power spectra~\cite{Planck:2018jri}.  This motivates further work on   similar   concrete models, where the  oscillation amplitude at CMB scales is reduced, while still attaining large oscillations and growth of $P_{\mathcal{R}}(k)$ at small scales. A particularly interesting outcome of this model is that it represents the first concrete example of resonant oscillatory feature in the power spectrum, which has been only explored phenomenologically (with oscillation templates) before~\cite{Gao:2015aba,Planck:2018jri,Fumagalli:2020adf,Fumagalli:2020nvq,Fumagalli:2021cel}. 
The resulting $P_{\mathcal{R}}(k)$ executes logarithmic oscillations with varying frequencies. For a model which is consistent with the CMB bounds at large scales and also exhibits peaked profile of resonant oscillations at small scales, it will be interesting to find a fit of $P_{\mathcal{R}}(k)$ in $\log (k)$ around the peak, which we would like to explore in the future. 

Interestingly, $P_{\mathcal{R}}(k)$ can be large enough for suitable choices of the model parameters to generate abundant PBHs. In Fig.~\ref{fig:PBHcomp}, we have shown the weighted mass spectra for PBH $M_{\rm PBH}\psi (M_{\rm PBH})$ for some chosen examples and obtained $\fpbh \sim 10^{-4} - 10^{-2}$ for light masses ($10^{-24} \ms- 10^{-16}\ms$), which have constraints from BBN and CMB. The presence of multiple peaks in the PBH mass spectra may be useful to generate abundant PBHs with a few target masses by tuning the model parameters. The PBH analysis in section~\ref{secPBH} is done only for the standard RD epoch. However, an extended reheating epoch or an additional scalar dominated epoch (e.o.s. $w_e$) may be present after inflation, particularly, since $N_{\rm inf}$ can be large to enforce CMB-scale red-tilt in some cases. Then, if the peak mode $k_p$ enters the post-inflationary horizon in this non-standard epoch, the dependence between $M_{\rm PBH}$ and $k$ changes~\cite{Domenech:2021ztg} since $M_{\rm PBH}\propto k_p^{-2-a}$, where $a=\frac{1-3w_e}{1+3w_e}$. Thus, the PBH produced in a $w_e$-dominated epoch will have a higher mass corresponding to the peak in the power spectrum at $k_p$ for $a<0$. This is possible when the universe is dominated by a stiff e.o.s. $w_e > 1/3$ after inflation and before BBN, and the PBH mass spectra (therefore $\fpbh$) are enhanced in this case~\cite{Bhattacharya:2019bvk}, which can be interesting in view of stringent limits on PBH abundance at low masses.

Enhancement in $P_{\mathcal{R}}(k)$ also leads to wide and large induced GW spectra as shown in Fig.~\ref{fig:GWcomp} for several fiducial set of parameters, which are eligible for prospective detection in multiple future GW surveys. Moreover, the characteristic 
resonant  oscillations of $P_{\mathcal{R}}(k)$ on top of the peak profile are also imprinted in the GW spectra, which can be verified and constrained in case of detection or overall upper bounds of $P_{\mathcal{R}}(k)$ can be derived (for non-detection), which can constrain $\lambda/M$ and $b$. 

In Table~\ref{tab:2}, we have quoted the tensor-to-scalar ratios $r$ for each set of parameters, which is within the latest CMB bound~\cite{BICEP:2021xfz}. However, next generation of CMB surveys such as LiteBird~\cite{2018cosp...42E1401H}, CMB-S4~\cite{CMB-S4:2016ple}, CORE etc.~propose to improve the bound on $r$ by a few orders, which can constrain or even refute the model under consideration.

Let us  stress that the parameter choices  in Table~\ref{tab:2}, which are used to show the profiles of $P_{\mathcal{R}}(k)$ (Fig.~\ref{fig:pscomp}), $M\psi (M)$ (Fig.~\ref{fig:PBHcomp}) and $\Omega _{\rm GW}$ (Fig.~\ref{fig:GWcomp}), represent concrete  examples to illustrate how the mechanism we introduced can be easily realised in  a compelling  supergravity model of axion monodromy and the rich phenomenology  that can arise from this class of models. 
However, as we indicated in the Introduction and discussed in Sec.~\ref{secPBH} and~\ref{secGW}, 
none of  these parameter combinations are fully consistent with  observational constraints when the phenomenology at all the scales is considered. For example, in the three cases for which the PBH mass spectra are presented in Fig.~\ref{fig:PBHcomp}, observational bounds on PBH abundance are possibly violated when exact bounds for the extended mass functions are evaluated. On the other hand, for $\lambda / M = 90$, $b=50$, light PBHs are produced with the abundance $\fpbh \sim 10^{-4}$ and the induced GWs have a large, wide spectrum which indicates to interesting and rich small-scale phenomenology. However, as can be seen from the left panel of Fig.~\ref{fig:pscomp}, for the examples in Table~\ref{tab:2} with $\lambda / M \geq 90$, despite the pivot-scale values of $n_s$ and $r$  
being in agreement with CMB constraints, the large scale $P_{\mathcal{R}}(k)$ has no distinct red-tilt. On the other hand, for the first 3 and last 2 examples in Table~\ref{tab:2}, overall red-tilt of the oscillatory envelope of $P_{\mathcal{R}}(k)$ at large scales can be seen in Fig.~\ref{fig:pscomp}.  
Of  these, the cases $\lambda / M =70, b=50$ and $\lambda / M =80, b=50$ lead to interesting small-scale phenomenology for PBH and/or induced GW in view of the expected sensitivities of future surveys. 
However, we emphasise again that due to large oscillations at the CMB scales, even the examples with $\lambda / M \leq 80, b=50$ are in tension with CMB observations. 
Moreover, there can be substantial non-Gaussianities resulting from such transient large turns, which need to be explored and may be instrumental to constrain the model parameters more stringently~\cite{Garcia-Saenz:2018vqf,Fumagalli:2019noh}.

The present model of axion monodromy in supergravity is one example of a class of models where the subleading non-perturbative corrections to the axion scalar potential  modify the potential to induce transient slow-roll violations. It would be interesting to investigate whether these corrections to the axion in K\"ahler or Fibre  inflation could give similar interesting phenomenologies. 

\newpage

\acknowledgments{We are  thankful to Suman Chatterjee for help and discussion about numerical codes. 
We further would like to thank Gianmssimo Tasinato for comments on the manuscript; Jacopo Fumagalli, Sebastian Renaux-Petel for discussions and comments on the first version of the paper. 
SB is supported by the ``Progetto di Eccellenza'' of the Department of Physics and Astronomy of the University of Padua. She also acknowledges support by Istituto Nazionale di Fisica Nucleare (INFN) through the Theoretical Astroparticle Physics (TAsP) project. SB thanks Science and Engineering Research Board, Government of India (SERB, GoI) for support during initial part of this work. IZ is partially supported by STFC, grant  ST/T000813/1. }

\bibliographystyle{utphys}
\bibliography{refsBZ2}

\providecommand{\href}[2]{#2}\begingroup\raggedright\begin{thebibliography}{100}

\bibitem{Planck:2018jri}
{\bf Planck} Collaboration, Y.~Akrami {\em et al.}, ``{Planck 2018 results. X.
  Constraints on inflation},''
  \href{http://dx.doi.org/10.1051/0004-6361/201833887}{{\em Astron. Astrophys.}
  {\bf 641} (2020)  A10}, \href{http://arxiv.org/abs/1807.06211}{{\tt
  arXiv:1807.06211 [astro-ph.CO]}}.

\bibitem{BICEP:2021xfz}
{\bf BICEP, Keck} Collaboration, P.~A.~R. Ade {\em et al.}, ``{Improved
  Constraints on Primordial Gravitational Waves using Planck, WMAP, and
  BICEP/Keck Observations through the 2018 Observing Season},''
  \href{http://dx.doi.org/10.1103/PhysRevLett.127.151301}{{\em Phys. Rev.
  Lett.} {\bf 127} (2021) no.~15, 151301},
  \href{http://arxiv.org/abs/2110.00483}{{\tt arXiv:2110.00483 [astro-ph.CO]}}.

\bibitem{Starobinsky:1992ts}
A.~A. Starobinsky, ``{Spectrum of adiabatic perturbations in the universe when
  there are singularities in the inflation potential},'' {\em JETP Lett.} {\bf
  55} (1992)  489--494.

\bibitem{Ivanov:1994pa}
P.~Ivanov, P.~Naselsky, and I.~Novikov, ``{Inflation and primordial black holes
  as dark matter},'' \href{http://dx.doi.org/10.1103/PhysRevD.50.7173}{{\em
  Phys. Rev. D} {\bf 50} (1994)  7173--7178}.

\bibitem{Garcia-Bellido:2017mdw}
J.~Garcia-Bellido and E.~Ruiz~Morales, ``{Primordial black holes from single
  field models of inflation},''
  \href{http://dx.doi.org/10.1016/j.dark.2017.09.007}{{\em Phys. Dark Univ.}
  {\bf 18} (2017)  47--54},
\href{http://arxiv.org/abs/1702.03901}{{\tt arXiv:1702.03901 [astro-ph.CO]}}.

\bibitem{Bhaumik:2019tvl}
N.~Bhaumik and R.~K. Jain, ``{Primordial black holes dark matter from
  inflection point models of inflation and the effects of reheating},''
  \href{http://dx.doi.org/10.1088/1475-7516/2020/01/037}{{\em JCAP} {\bf 01}
  (2020)  037}, \href{http://arxiv.org/abs/1907.04125}{{\tt arXiv:1907.04125
  [astro-ph.CO]}}.

\bibitem{Mishra:2019pzq}
S.~S. Mishra and V.~Sahni, ``{Primordial Black Holes from a tiny bump/dip in
  the Inflaton potential},''
  \href{http://dx.doi.org/10.1088/1475-7516/2020/04/007}{{\em JCAP} {\bf 04}
  (2020)  007}, \href{http://arxiv.org/abs/1911.00057}{{\tt arXiv:1911.00057
  [gr-qc]}}.

\bibitem{RPT}
S.~Renaux-Petel and K.~Turzynski, ``{Geometrical Destabilization of
  Inflation},'' \href{http://dx.doi.org/10.1103/PhysRevLett.117.141301}{{\em
  Phys. Rev. Lett.} {\bf 117} (2016) no.~14, 141301},
\href{http://arxiv.org/abs/1510.01281}{{\tt arXiv:1510.01281 [astro-ph.CO]}}.

\bibitem{Brown:2017osf}
A.~R. Brown, ``{Hyperbolic Inflation},''
  \href{http://dx.doi.org/10.1103/PhysRevLett.121.251601}{{\em Phys. Rev.
  Lett.} {\bf 121} (2018) no.~25, 251601},
  \href{http://arxiv.org/abs/1705.03023}{{\tt arXiv:1705.03023 [hep-th]}}.

\bibitem{Christodoulidis:2018qdw}
P.~Christodoulidis, D.~Roest, and E.~I. Sfakianakis, ``{Angular inflation in
  multi-field $\alpha$-attractors},''
  \href{http://dx.doi.org/10.1088/1475-7516/2019/11/002}{{\em JCAP} {\bf 11}
  (2019)  002}, \href{http://arxiv.org/abs/1803.09841}{{\tt arXiv:1803.09841
  [hep-th]}}.

\bibitem{Garcia-Saenz:2018ifx}
S.~Garcia-Saenz, S.~Renaux-Petel, and J.~Ronayne, ``{Primordial fluctuations
  and non-Gaussianities in sidetracked inflation},''
  \href{http://dx.doi.org/10.1088/1475-7516/2018/07/057}{{\em JCAP} {\bf 07}
  (2018)  057}, \href{http://arxiv.org/abs/1804.11279}{{\tt arXiv:1804.11279
  [astro-ph.CO]}}.

\bibitem{Garcia-Saenz:2018vqf}
S.~Garcia-Saenz and S.~Renaux-Petel, ``{Flattened non-Gaussianities from the
  effective field theory of inflation with imaginary speed of sound},''
  \href{http://dx.doi.org/10.1088/1475-7516/2018/11/005}{{\em JCAP} {\bf 11}
  (2018)  005}, \href{http://arxiv.org/abs/1805.12563}{{\tt arXiv:1805.12563
  [hep-th]}}.

\bibitem{Bjorkmo:2019aev}
T.~Bjorkmo and M.~D. Marsh, ``{Hyperinflation generalised: from its attractor
  mechanism to its tension with the \textquoteleft{}swampland
  conditions\textquoteright{}},''
  \href{http://dx.doi.org/10.1007/JHEP04(2019)172}{{\em JHEP} {\bf 04} (2019)
  172}, \href{http://arxiv.org/abs/1901.08603}{{\tt arXiv:1901.08603
  [hep-th]}}.

\bibitem{Bjorkmo:2019fls}
T.~Bjorkmo, ``{Rapid-Turn Inflationary Attractors},''
  \href{http://dx.doi.org/10.1103/PhysRevLett.122.251301}{{\em Phys. Rev.
  Lett.} {\bf 122} (2019) no.~25, 251301},
  \href{http://arxiv.org/abs/1902.10529}{{\tt arXiv:1902.10529 [hep-th]}}.

\bibitem{Fumagalli:2019noh}
J.~Fumagalli, S.~Garcia-Saenz, L.~Pinol, S.~Renaux-Petel, and J.~Ronayne,
  ``{Hyper-Non-Gaussianities in Inflation with Strongly Nongeodesic Motion},''
  \href{http://dx.doi.org/10.1103/PhysRevLett.123.201302}{{\em Phys. Rev.
  Lett.} {\bf 123} (2019) no.~20, 201302},
  \href{http://arxiv.org/abs/1902.03221}{{\tt arXiv:1902.03221 [hep-th]}}.

\bibitem{Christodoulidis:2019mkj}
P.~Christodoulidis, D.~Roest, and E.~I. Sfakianakis, ``{Attractors,
  Bifurcations and Curvature in Multi-field Inflation},''
  \href{http://dx.doi.org/10.1088/1475-7516/2020/08/006}{{\em JCAP} {\bf 08}
  (2020)  006}, \href{http://arxiv.org/abs/1903.03513}{{\tt arXiv:1903.03513
  [gr-qc]}}.

\bibitem{Christodoulidis:2019jsx}
P.~Christodoulidis, D.~Roest, and E.~I. Sfakianakis, ``{Scaling attractors in
  multi-field inflation},''
  \href{http://dx.doi.org/10.1088/1475-7516/2019/12/059}{{\em JCAP} {\bf 12}
  (2019)  059}, \href{http://arxiv.org/abs/1903.06116}{{\tt arXiv:1903.06116
  [hep-th]}}.

\bibitem{Aragam:2020uqi}
V.~Aragam, S.~Paban, and R.~Rosati, ``{The Multi-Field, Rapid-Turn Inflationary
  Solution},'' \href{http://arxiv.org/abs/2010.15933}{{\tt arXiv:2010.15933
  [hep-th]}}.

\bibitem{Aragam:2021scu}
V.~Aragam, R.~Chiovoloni, S.~Paban, R.~Rosati, and I.~Zavala, ``{Rapid-turn
  inflation in supergravity is rare and tachyonic},''
  \href{http://arxiv.org/abs/2110.05516}{{\tt arXiv:2110.05516 [hep-th]}}.

\bibitem{Chakraborty:2019dfh}
D.~Chakraborty, R.~Chiovoloni, O.~Loaiza-Brito, G.~Niz, and I.~Zavala, ``{Fat
  inflatons, large turns and the $\eta$-problem},''
  \href{http://dx.doi.org/10.1088/1475-7516/2020/01/020}{{\em JCAP} {\bf 01}
  (2020)  020}, \href{http://arxiv.org/abs/1908.09797}{{\tt arXiv:1908.09797
  [hep-th]}}.

\bibitem{Flauger:2009ab}
R.~Flauger, L.~McAllister, E.~Pajer, A.~Westphal, and G.~Xu, ``{Oscillations in
  the CMB from Axion Monodromy Inflation},''
  \href{http://dx.doi.org/10.1088/1475-7516/2010/06/009}{{\em JCAP} {\bf 06}
  (2010)  009}, \href{http://arxiv.org/abs/0907.2916}{{\tt arXiv:0907.2916
  [hep-th]}}.

\bibitem{Flauger:2014ana}
R.~Flauger, L.~McAllister, E.~Silverstein, and A.~Westphal, ``{Drifting
  Oscillations in Axion Monodromy},''
  \href{http://dx.doi.org/10.1088/1475-7516/2017/10/055}{{\em JCAP} {\bf 10}
  (2017)  055}, \href{http://arxiv.org/abs/1412.1814}{{\tt arXiv:1412.1814
  [hep-th]}}.

\bibitem{Gao:2015aba}
X.~Gao and J.-O. Gong, ``{Towards general patterns of features in multi-field
  inflation},'' \href{http://dx.doi.org/10.1007/JHEP08(2015)115}{{\em JHEP}
  {\bf 08} (2015)  115}, \href{http://arxiv.org/abs/1506.08894}{{\tt
  arXiv:1506.08894 [astro-ph.CO]}}.

\bibitem{Fumagalli:2020adf}
J.~Fumagalli, S.~Renaux-Petel, J.~W. Ronayne, and L.~T. Witkowski, ``{Turning
  in the landscape: a new mechanism for generating Primordial Black Holes},''
  \href{http://arxiv.org/abs/2004.08369}{{\tt arXiv:2004.08369 [hep-th]}}.

\bibitem{Fumagalli:2020nvq}
J.~Fumagalli, S.~Renaux-Petel, and L.~T. Witkowski, ``{Oscillations in the
  stochastic gravitational wave background from sharp features and particle
  production during inflation},''
  \href{http://dx.doi.org/10.1088/1475-7516/2021/08/030}{{\em JCAP} {\bf 08}
  (2021)  030}, \href{http://arxiv.org/abs/2012.02761}{{\tt arXiv:2012.02761
  [astro-ph.CO]}}.

\bibitem{Braglia:2020taf}
M.~Braglia, X.~Chen, and D.~K. Hazra, ``{Probing Primordial Features with the
  Stochastic Gravitational Wave Background},''
  \href{http://arxiv.org/abs/2012.05821}{{\tt arXiv:2012.05821 [astro-ph.CO]}}.

\bibitem{Fumagalli:2021cel}
J.~Fumagalli, S.~e. Renaux-Petel, and L.~T. Witkowski, ``{Resonant features in
  the stochastic gravitational wave background},''
  \href{http://dx.doi.org/10.1088/1475-7516/2021/08/059}{{\em JCAP} {\bf 08}
  (2021)  059}, \href{http://arxiv.org/abs/2105.06481}{{\tt arXiv:2105.06481
  [astro-ph.CO]}}.

\bibitem{Fumagalli:2021dtd}
J.~Fumagalli, M.~Pieroni, S.~Renaux-Petel, and L.~T. Witkowski, ``{Detecting
  primordial features with LISA},''
  \href{http://dx.doi.org/10.1088/1475-7516/2022/07/020}{{\em JCAP} {\bf 07}
  (2022) no.~07, 020}, \href{http://arxiv.org/abs/2112.06903}{{\tt
  arXiv:2112.06903 [astro-ph.CO]}}.

\bibitem{Carr:1974nx}
B.~J. Carr and S.~Hawking, ``{Black holes in the early Universe},'' {\em Mon.
  Not. Roy. Astron. Soc.} {\bf 168} (1974)  399--415.

\bibitem{Carr:1975qj}
B.~J. Carr, ``{The Primordial black hole mass spectrum},''
  \href{http://dx.doi.org/10.1086/153853}{{\em Astrophys. J.} {\bf 201} (1975)
  1--19}.

\bibitem{Carr:2021bzv}
B.~Carr and F.~Kuhnel, ``{Primordial Black Holes as Dark Matter Candidates},''
  \href{http://arxiv.org/abs/2110.02821}{{\tt arXiv:2110.02821 [astro-ph.CO]}}.

\bibitem{Hawking:1971ei}
S.~Hawking, ``{Gravitationally collapsed objects of very low mass},'' {\em Mon.
  Not. Roy. Astron. Soc.} {\bf 152} (1971)  75.

\bibitem{Hawking:1974rv}
S.~Hawking, ``{Black hole explosions},''
  \href{http://dx.doi.org/10.1038/248030a0}{{\em Nature} {\bf 248} (1974)
  30--31}.

\bibitem{Garcia-Bellido:1996mdl}
J.~Garcia-Bellido, A.~D. Linde, and D.~Wands, ``{Density perturbations and
  black hole formation in hybrid inflation},''
  \href{http://dx.doi.org/10.1103/PhysRevD.54.6040}{{\em Phys. Rev. D} {\bf 54}
  (1996)  6040--6058}, \href{http://arxiv.org/abs/astro-ph/9605094}{{\tt
  arXiv:astro-ph/9605094}}.

\bibitem{Aldabergenov:2020yok}
Y.~Aldabergenov, A.~Addazi, and S.~V. Ketov, ``{Testing Primordial Black Holes
  as Dark Matter in Supergravity from Gravitational Waves},''
  \href{http://dx.doi.org/10.1016/j.physletb.2021.136069}{{\em Phys. Lett. B}
  {\bf 814} (2021)  136069}, \href{http://arxiv.org/abs/2008.10476}{{\tt
  arXiv:2008.10476 [hep-th]}}.

\bibitem{Aldabergenov:2020bpt}
Y.~Aldabergenov, A.~Addazi, and S.~V. Ketov, ``{Primordial black holes from
  modified supergravity},''
  \href{http://dx.doi.org/10.1140/epjc/s10052-020-08506-6}{{\em Eur. Phys. J.
  C} {\bf 80} (2020) no.~10, 917}, \href{http://arxiv.org/abs/2006.16641}{{\tt
  arXiv:2006.16641 [hep-th]}}.

\bibitem{Ishikawa:2021xya}
R.~Ishikawa and S.~V. Ketov, ``{Exploring the parameter space of modified
  supergravity for double inflation and primordial black hole formation},''
  \href{http://dx.doi.org/10.1088/1361-6382/ac3bd9}{{\em Class. Quant. Grav.}
  {\bf 39} (2022) no.~1, 015016}, \href{http://arxiv.org/abs/2108.04408}{{\tt
  arXiv:2108.04408 [astro-ph.CO]}}.

\bibitem{Ketov:2021fww}
S.~V. Ketov, ``{Multi-Field versus Single-Field in the Supergravity Models of
  Inflation and Primordial Black Holes},''
  \href{http://dx.doi.org/10.3390/universe7050115}{{\em Universe} {\bf 7}
  (2021) no.~5, 115}.

\bibitem{Pi:2017gih}
S.~Pi, Y.-l. Zhang, Q.-G. Huang, and M.~Sasaki, ``{Scalaron from $R^2$-gravity
  as a heavy field},''
  \href{http://dx.doi.org/10.1088/1475-7516/2018/05/042}{{\em JCAP} {\bf 05}
  (2018)  042}, \href{http://arxiv.org/abs/1712.09896}{{\tt arXiv:1712.09896
  [astro-ph.CO]}}.

\bibitem{Braglia:2020eai}
M.~Braglia, D.~K. Hazra, F.~Finelli, G.~F. Smoot, L.~Sriramkumar, and A.~A.
  Starobinsky, ``{Generating PBHs and small-scale GWs in two-field models of
  inflation},'' \href{http://dx.doi.org/10.1088/1475-7516/2020/08/001}{{\em
  JCAP} {\bf 08} (2020)  001}, \href{http://arxiv.org/abs/2005.02895}{{\tt
  arXiv:2005.02895 [astro-ph.CO]}}.

\bibitem{Anguelova:2020nzl}
L.~Anguelova, ``{On Primordial Black Holes from Rapid Turns in Two-field
  Models},'' \href{http://dx.doi.org/10.1088/1475-7516/2021/06/004}{{\em JCAP}
  {\bf 06} (2021)  004}, \href{http://arxiv.org/abs/2012.03705}{{\tt
  arXiv:2012.03705 [hep-th]}}.

\bibitem{Palma:2020ejf}
G.~A. Palma, S.~Sypsas, and C.~Zenteno, ``{Seeding primordial black holes in
  multifield inflation},''
  \href{http://dx.doi.org/10.1103/PhysRevLett.125.121301}{{\em Phys. Rev.
  Lett.} {\bf 125} (2020) no.~12, 121301},
  \href{http://arxiv.org/abs/2004.06106}{{\tt arXiv:2004.06106 [astro-ph.CO]}}.

\bibitem{Ananda:2006af}
K.~N. Ananda, C.~Clarkson, and D.~Wands, ``{The Cosmological gravitational wave
  background from primordial density perturbations},''
  \href{http://dx.doi.org/10.1103/PhysRevD.75.123518}{{\em Phys. Rev. D} {\bf
  75} (2007)  123518}, \href{http://arxiv.org/abs/gr-qc/0612013}{{\tt
  arXiv:gr-qc/0612013}}.

\bibitem{Baumann:2007zm}
D.~Baumann, P.~J. Steinhardt, K.~Takahashi, and K.~Ichiki, ``{Gravitational
  Wave Spectrum Induced by Primordial Scalar Perturbations},''
  \href{http://dx.doi.org/10.1103/PhysRevD.76.084019}{{\em Phys. Rev. D} {\bf
  76} (2007)  084019}, \href{http://arxiv.org/abs/hep-th/0703290}{{\tt
  arXiv:hep-th/0703290}}.

\bibitem{Domenech:2021ztg}
G.~Dom\`enech, ``{Scalar Induced Gravitational Waves Review},''
  \href{http://dx.doi.org/10.3390/universe7110398}{{\em Universe} {\bf 7}
  (2021) no.~11, 398}, \href{http://arxiv.org/abs/2109.01398}{{\tt
  arXiv:2109.01398 [gr-qc]}}.

\bibitem{Witkowski:2021raz}
L.~T. Witkowski, G.~Dom\`enech, J.~Fumagalli, and S.~Renaux-Petel, ``{Expansion
  history-dependent oscillations in the scalar-induced gravitational wave
  background},'' \href{http://arxiv.org/abs/2110.09480}{{\tt arXiv:2110.09480
  [astro-ph.CO]}}.

\bibitem{Fumagalli:2021mpc}
J.~Fumagalli, G.~A. Palma, S.~Renaux-Petel, S.~Sypsas, L.~T. Witkowski, and
  C.~Zenteno, ``{Primordial gravitational waves from excited states},''
  \href{http://dx.doi.org/10.1007/JHEP03(2022)196}{{\em JHEP} {\bf 03} (2022)
  196}, \href{http://arxiv.org/abs/2111.14664}{{\tt arXiv:2111.14664
  [astro-ph.CO]}}.

\bibitem{EGNO}
J.~Ellis, M.~A.~G. Garcia, D.~V. Nanopoulos, and K.~A. Olive, ``{A No-Scale
  Inflationary Model to Fit Them All},''
  \href{http://dx.doi.org/10.1088/1475-7516/2014/08/044}{{\em JCAP} {\bf 08}
  (2014)  044}, \href{http://arxiv.org/abs/1405.0271}{{\tt arXiv:1405.0271
  [hep-ph]}}.

\bibitem{Kobayashi:2015aaa}
T.~Kobayashi, A.~Oikawa, and H.~Otsuka, ``{New potentials for string axion
  inflation},'' \href{http://dx.doi.org/10.1103/PhysRevD.93.083508}{{\em Phys.
  Rev. D} {\bf 93} (2016) no.~8, 083508},
  \href{http://arxiv.org/abs/1510.08768}{{\tt arXiv:1510.08768 [hep-ph]}}.

\bibitem{CaboBizet:2016uzv}
N.~Cabo~Bizet, O.~Loaiza-Brito, and I.~Zavala, ``{Mirror quintic vacua:
  hierarchies and inflation},''
  \href{http://dx.doi.org/10.1007/JHEP10(2016)082}{{\em JHEP} {\bf 10} (2016)
  082}, \href{http://arxiv.org/abs/1605.03974}{{\tt arXiv:1605.03974
  [hep-th]}}.

\bibitem{Parameswaran:2016qqq}
S.~Parameswaran, G.~Tasinato, and I.~Zavala, ``{Subleading Effects and the
  Field Range in Axion Inflation},''
  \href{http://dx.doi.org/10.1088/1475-7516/2016/04/008}{{\em JCAP} {\bf 04}
  (2016)  008}, \href{http://arxiv.org/abs/1602.02812}{{\tt arXiv:1602.02812
  [astro-ph.CO]}}.

\bibitem{Ozsoy:2018flq}
O.~\"Ozsoy, S.~Parameswaran, G.~Tasinato, and I.~Zavala, ``{Mechanisms for
  Primordial Black Hole Production in String Theory},''
  \href{http://dx.doi.org/10.1088/1475-7516/2018/07/005}{{\em JCAP} {\bf 07}
  (2018)  005}, \href{http://arxiv.org/abs/1803.07626}{{\tt arXiv:1803.07626
  [hep-th]}}.

\bibitem{DAmico:2020euu}
G.~D'Amico and N.~Kaloper, ``{Rollercoaster cosmology},''
  \href{http://dx.doi.org/10.1088/1475-7516/2021/08/058}{{\em JCAP} {\bf 08}
  (2021)  058}, \href{http://arxiv.org/abs/2011.09489}{{\tt arXiv:2011.09489
  [hep-th]}}.

\bibitem{DAmico:2021vka}
G.~D'Amico, N.~Kaloper, and A.~Westphal, ``{Double Monodromy Inflation: A
  Gravity Waves Factory for CMB-S4, LiteBIRD and LISA},''
  \href{http://dx.doi.org/10.1103/PhysRevD.104.L081302}{{\em Phys. Rev. D} {\bf
  104} (2021) no.~8, L081302}, \href{http://arxiv.org/abs/2101.05861}{{\tt
  arXiv:2101.05861 [hep-th]}}.

\bibitem{DAmico:2021fhz}
G.~D'Amico, N.~Kaloper, and A.~Westphal, ``{General double monodromy
  inflation},'' \href{http://dx.doi.org/10.1103/PhysRevD.105.103527}{{\em Phys.
  Rev. D} {\bf 105} (2022) no.~10, 103527},
  \href{http://arxiv.org/abs/2112.13861}{{\tt arXiv:2112.13861 [hep-th]}}.

\bibitem{AM1}
L.~McAllister, E.~Silverstein, and A.~Westphal, ``{Gravity Waves and Linear
  Inflation from Axion Monodromy},''
  \href{http://dx.doi.org/10.1103/PhysRevD.82.046003}{{\em Phys. Rev. D} {\bf
  82} (2010)  046003}, \href{http://arxiv.org/abs/0808.0706}{{\tt
  arXiv:0808.0706 [hep-th]}}.

\bibitem{AM2}
L.~McAllister, E.~Silverstein, A.~Westphal, and T.~Wrase, ``{The Powers of
  Monodromy},'' \href{http://dx.doi.org/10.1007/JHEP09(2014)123}{{\em JHEP}
  {\bf 09} (2014)  123}, \href{http://arxiv.org/abs/1405.3652}{{\tt
  arXiv:1405.3652 [hep-th]}}.

\bibitem{Achucarro:2010da}
A.~Achucarro, J.-O. Gong, S.~Hardeman, G.~A. Palma, and S.~P. Patil,
  ``{Features of heavy physics in the CMB power spectrum},''
  \href{http://dx.doi.org/10.1088/1475-7516/2011/01/030}{{\em JCAP} {\bf 01}
  (2011)  030}, \href{http://arxiv.org/abs/1010.3693}{{\tt arXiv:1010.3693
  [hep-ph]}}.

\bibitem{Hetz:2016ics}
A.~Hetz and G.~A. Palma, ``{Sound Speed of Primordial Fluctuations in
  Supergravity Inflation},''
  \href{http://dx.doi.org/10.1103/PhysRevLett.117.101301}{{\em Phys. Rev.
  Lett.} {\bf 117} (2016) no.~10, 101301},
  \href{http://arxiv.org/abs/1601.05457}{{\tt arXiv:1601.05457 [hep-th]}}.

\bibitem{Garg:2018reu}
S.~K. Garg and C.~Krishnan, ``{Bounds on Slow Roll and the de Sitter
  Swampland},'' \href{http://dx.doi.org/10.1007/JHEP11(2019)075}{{\em JHEP}
  {\bf 11} (2019)  075}, \href{http://arxiv.org/abs/1807.05193}{{\tt
  arXiv:1807.05193 [hep-th]}}.

\bibitem{Boutivas:2022qtl}
K.~Boutivas, I.~Dalianis, G.~P. Kodaxis, and N.~Tetradis, ``{The effect of
  multiple features on the power spectrum in two-field inflation},''
  \href{http://arxiv.org/abs/2203.15605}{{\tt arXiv:2203.15605 [astro-ph.CO]}}.

\bibitem{Kawasaki:2000yn}
M.~Kawasaki, M.~Yamaguchi, and T.~Yanagida, ``{Natural chaotic inflation in
  supergravity},'' \href{http://dx.doi.org/10.1103/PhysRevLett.85.3572}{{\em
  Phys. Rev. Lett.} {\bf 85} (2000)  3572--3575},
  \href{http://arxiv.org/abs/hep-ph/0004243}{{\tt arXiv:hep-ph/0004243}}.

\bibitem{KLR}
R.~Kallosh, A.~Linde, and T.~Rube, ``{General inflaton potentials in
  supergravity},'' \href{http://dx.doi.org/10.1103/PhysRevD.83.043507}{{\em
  Phys. Rev. D} {\bf 83} (2011)  043507},
  \href{http://arxiv.org/abs/1011.5945}{{\tt arXiv:1011.5945 [hep-th]}}.

\bibitem{Ferrara:2014kva}
S.~Ferrara, R.~Kallosh, and A.~Linde, ``{Cosmology with Nilpotent
  Superfields},'' \href{http://dx.doi.org/10.1007/JHEP10(2014)143}{{\em JHEP}
  {\bf 10} (2014)  143}, \href{http://arxiv.org/abs/1408.4096}{{\tt
  arXiv:1408.4096 [hep-th]}}.

\bibitem{Roest:2013aoa}
D.~Roest, M.~Scalisi, and I.~Zavala, ``{K\"ahler potentials for Planck
  inflation},'' \href{http://dx.doi.org/10.1088/1475-7516/2013/11/007}{{\em
  JCAP} {\bf 11} (2013)  007}, \href{http://arxiv.org/abs/1307.4343}{{\tt
  arXiv:1307.4343 [hep-th]}}.

\bibitem{WGC}
N.~Arkani-Hamed, L.~Motl, A.~Nicolis, and C.~Vafa, ``{The String landscape,
  black holes and gravity as the weakest force},''
  \href{http://dx.doi.org/10.1088/1126-6708/2007/06/060}{{\em JHEP} {\bf 06}
  (2007)  060}, \href{http://arxiv.org/abs/hep-th/0601001}{{\tt
  arXiv:hep-th/0601001}}.

\bibitem{SS}
M.~Sasaki and E.~D. Stewart, ``{A General analytic formula for the spectral
  index of the density perturbations produced during inflation},''
  \href{http://dx.doi.org/10.1143/PTP.95.71}{{\em Prog. Theor. Phys.} {\bf 95}
  (1996)  71--78},
\href{http://arxiv.org/abs/astro-ph/9507001}{{\tt arXiv:astro-ph/9507001
  [astro-ph]}}.

\bibitem{GWBM}
C.~Gordon, D.~Wands, B.~A. Bassett, and R.~Maartens, ``{Adiabatic and entropy
  perturbations from inflation},''
  \href{http://dx.doi.org/10.1103/PhysRevD.63.023506}{{\em Phys. Rev.} {\bf
  D63} (2001)  023506},
\href{http://arxiv.org/abs/astro-ph/0009131}{{\tt arXiv:astro-ph/0009131
  [astro-ph]}}.

\bibitem{GNvT}
S.~Groot~Nibbelink and B.~J.~W. van Tent, ``{Scalar perturbations during
  multiple field slow-roll inflation},''
  \href{http://dx.doi.org/10.1088/0264-9381/19/4/302}{{\em Class. Quant. Grav.}
  {\bf 19} (2002)  613--640},
\href{http://arxiv.org/abs/hep-ph/0107272}{{\tt arXiv:hep-ph/0107272
  [hep-ph]}}.

\bibitem{LRP}
D.~Langlois and S.~Renaux-Petel, ``{Perturbations in generalized multi-field
  inflation},'' \href{http://dx.doi.org/10.1088/1475-7516/2008/04/017}{{\em
  JCAP} {\bf 0804} (2008)  017},
\href{http://arxiv.org/abs/0801.1085}{{\tt arXiv:0801.1085 [hep-th]}}.

\bibitem{Dimopoulos:2017ged}
K.~Dimopoulos, ``{Ultra slow-roll inflation demystified},''
  \href{http://dx.doi.org/10.1016/j.physletb.2017.10.066}{{\em Phys. Lett. B}
  {\bf 775} (2017)  262--265}, \href{http://arxiv.org/abs/1707.05644}{{\tt
  arXiv:1707.05644 [hep-ph]}}.

\bibitem{Kinney:2005vj}
W.~H. Kinney, ``{Horizon crossing and inflation with large eta},''
  \href{http://dx.doi.org/10.1103/PhysRevD.72.023515}{{\em Phys. Rev. D} {\bf
  72} (2005)  023515}, \href{http://arxiv.org/abs/gr-qc/0503017}{{\tt
  arXiv:gr-qc/0503017}}.

\bibitem{Hooshangi:2022lao}
S.~Hooshangi, A.~Talebian, M.~H. Namjoo, and H.~Firouzjahi, ``{Multiple field
  ultraslow-roll inflation: Primordial black holes from straight bulk and
  distorted boundary},''
  \href{http://dx.doi.org/10.1103/PhysRevD.105.083525}{{\em Phys. Rev. D} {\bf
  105} (2022) no.~8, 083525}, \href{http://arxiv.org/abs/2201.07258}{{\tt
  arXiv:2201.07258 [astro-ph.CO]}}.

\bibitem{Achucarro:2010jv}
A.~Achucarro, J.-O. Gong, S.~Hardeman, G.~A. Palma, and S.~P. Patil, ``{Mass
  hierarchies and non-decoupling in multi-scalar field dynamics},''
  \href{http://dx.doi.org/10.1103/PhysRevD.84.043502}{{\em Phys. Rev. D} {\bf
  84} (2011)  043502}, \href{http://arxiv.org/abs/1005.3848}{{\tt
  arXiv:1005.3848 [hep-th]}}.

\bibitem{Chen:2011zf}
X.~Chen, ``{Primordial Features as Evidence for Inflation},''
  \href{http://dx.doi.org/10.1088/1475-7516/2012/01/038}{{\em JCAP} {\bf 01}
  (2012)  038}, \href{http://arxiv.org/abs/1104.1323}{{\tt arXiv:1104.1323
  [hep-th]}}.

\bibitem{Shiu:2011qw}
G.~Shiu and J.~Xu, ``{Effective Field Theory and Decoupling in Multi-field
  Inflation: An Illustrative Case Study},''
  \href{http://dx.doi.org/10.1103/PhysRevD.84.103509}{{\em Phys. Rev. D} {\bf
  84} (2011)  103509}, \href{http://arxiv.org/abs/1108.0981}{{\tt
  arXiv:1108.0981 [hep-th]}}.

\bibitem{Cespedes:2012hu}
S.~Cespedes, V.~Atal, and G.~A. Palma, ``{On the importance of heavy fields
  during inflation},''
  \href{http://dx.doi.org/10.1088/1475-7516/2012/05/008}{{\em JCAP} {\bf 05}
  (2012)  008}, \href{http://arxiv.org/abs/1201.4848}{{\tt arXiv:1201.4848
  [hep-th]}}.

\bibitem{Achucarro:2012sm}
A.~Achucarro, J.-O. Gong, S.~Hardeman, G.~A. Palma, and S.~P. Patil,
  ``{Effective theories of single field inflation when heavy fields matter},''
  \href{http://dx.doi.org/10.1007/JHEP05(2012)066}{{\em JHEP} {\bf 05} (2012)
  066}, \href{http://arxiv.org/abs/1201.6342}{{\tt arXiv:1201.6342 [hep-th]}}.

\bibitem{Avgoustidis:2012yc}
A.~Avgoustidis, S.~Cremonini, A.-C. Davis, R.~H. Ribeiro, K.~Turzynski, and
  S.~Watson, ``{Decoupling Survives Inflation: A Critical Look at Effective
  Field Theory Violations During Inflation},''
  \href{http://dx.doi.org/10.1088/1475-7516/2012/06/025}{{\em JCAP} {\bf 06}
  (2012)  025}, \href{http://arxiv.org/abs/1203.0016}{{\tt arXiv:1203.0016
  [hep-th]}}.

\bibitem{Gao:2012uq}
X.~Gao, D.~Langlois, and S.~Mizuno, ``{Influence of heavy modes on
  perturbations in multiple field inflation},''
  \href{http://dx.doi.org/10.1088/1475-7516/2012/10/040}{{\em JCAP} {\bf 10}
  (2012)  040}, \href{http://arxiv.org/abs/1205.5275}{{\tt arXiv:1205.5275
  [hep-th]}}.

\bibitem{Achucarro:2012fd}
A.~Ach\'ucarro, J.-O. Gong, G.~A. Palma, and S.~P. Patil, ``{Correlating
  features in the primordial spectra},''
  \href{http://dx.doi.org/10.1103/PhysRevD.87.121301}{{\em Phys. Rev. D} {\bf
  87} (2013) no.~12, 121301}, \href{http://arxiv.org/abs/1211.5619}{{\tt
  arXiv:1211.5619 [astro-ph.CO]}}.

\bibitem{Konieczka:2014zja}
M.~Konieczka, R.~H. Ribeiro, and K.~Turzynski, ``{The effects of a fast-turning
  trajectory in multiple-field inflation},''
  \href{http://dx.doi.org/10.1088/1475-7516/2014/07/030}{{\em JCAP} {\bf 07}
  (2014)  030}, \href{http://arxiv.org/abs/1401.6163}{{\tt arXiv:1401.6163
  [astro-ph.CO]}}.

\bibitem{Iacconi:2021ltm}
L.~Iacconi, H.~Assadullahi, M.~Fasiello, and D.~Wands, ``{Revisiting
  small-scale fluctuations in $\alpha$-attractor models of inflation},''
  \href{http://arxiv.org/abs/2112.05092}{{\tt arXiv:2112.05092 [astro-ph.CO]}}.

\bibitem{Mulryne:2016mzv}
D.~J. Mulryne and J.~W. Ronayne, ``{PyTransport: A Python package for the
  calculation of inflationary correlation functions},''
  \href{http://dx.doi.org/10.21105/joss.00494}{{\em J. Open Source Softw.} {\bf
  3} (2018) no.~23, 494}, \href{http://arxiv.org/abs/1609.00381}{{\tt
  arXiv:1609.00381 [astro-ph.CO]}}.

\bibitem{Braglia:2021rej}
M.~Braglia, X.~Chen, and D.~K. Hazra, ``{Primordial standard clock models and
  CMB residual anomalies},''
  \href{http://dx.doi.org/10.1103/PhysRevD.105.103523}{{\em Phys. Rev. D} {\bf
  105} (2022) no.~10, 103523}, \href{http://arxiv.org/abs/2108.10110}{{\tt
  arXiv:2108.10110 [astro-ph.CO]}}.

\bibitem{Adshead:2010mc}
P.~Adshead, R.~Easther, J.~Pritchard, and A.~Loeb, ``{Inflation and the Scale
  Dependent Spectral Index: Prospects and Strategies},''
  \href{http://dx.doi.org/10.1088/1475-7516/2011/02/021}{{\em JCAP} {\bf 02}
  (2011)  021}, \href{http://arxiv.org/abs/1007.3748}{{\tt arXiv:1007.3748
  [astro-ph.CO]}}.

\bibitem{Maharana:2017fui}
A.~Maharana and I.~Zavala, ``{Postinflationary scalar tensor cosmology and
  inflationary parameters},''
  \href{http://dx.doi.org/10.1103/PhysRevD.97.123518}{{\em Phys. Rev. D} {\bf
  97} (2018) no.~12, 123518}, \href{http://arxiv.org/abs/1712.07071}{{\tt
  arXiv:1712.07071 [hep-ph]}}.

\bibitem{Churazov:2010wy}
E.~Churazov, S.~Sazonov, S.~Tsygankov, R.~Sunyaev, and D.~Varshalovich,
  ``{Positron annihilation spectrum from the Galactic Centre region observed by
  SPI/INTEGRAL, revisited: annihilation in a cooling ISM?},''
  \href{http://dx.doi.org/10.1111/j.1365-2966.2010.17804.x}{{\em Mon. Not. Roy.
  Astron. Soc.} {\bf 411} (2011)  1727},
  \href{http://arxiv.org/abs/1010.0864}{{\tt arXiv:1010.0864 [astro-ph.HE]}}.

\bibitem{Siegert:2016ijv}
T.~Siegert, R.~Diehl, A.~C. Vincent, F.~Guglielmetti, M.~G. Krause, and
  C.~Boehm, ``{Search for 511 keV Emission in Satellite Galaxies of the Milky
  Way with INTEGRAL/SPI},''
  \href{http://dx.doi.org/10.1051/0004-6361/201629136}{{\em Astron. Astrophys.}
  {\bf 595} (2016)  A25}, \href{http://arxiv.org/abs/1608.00393}{{\tt
  arXiv:1608.00393 [astro-ph.HE]}}.

\bibitem{Laha:2019ssq}
R.~Laha, ``{Primordial Black Holes as a Dark Matter Candidate Are Severely
  Constrained by the Galactic Center 511 keV $\gamma$ -Ray Line},''
  \href{http://dx.doi.org/10.1103/PhysRevLett.123.251101}{{\em Phys. Rev.
  Lett.} {\bf 123} (2019) no.~25, 251101},
  \href{http://arxiv.org/abs/1906.09994}{{\tt arXiv:1906.09994 [astro-ph.HE]}}.

\bibitem{Bays:2011si}
{\bf Super-Kamiokande} Collaboration, K.~Bays {\em et al.}, ``{Supernova Relic
  Neutrino Search at Super-Kamiokande},''
  \href{http://dx.doi.org/10.1103/PhysRevD.85.052007}{{\em Phys. Rev. D} {\bf
  85} (2012)  052007}, \href{http://arxiv.org/abs/1111.5031}{{\tt
  arXiv:1111.5031 [hep-ex]}}.

\bibitem{Collaboration:2011jza}
{\bf KamLAND} Collaboration, A.~Gando {\em et al.}, ``{A study of
  extraterrestrial antineutrino sources with the KamLAND detector},''
  \href{http://dx.doi.org/10.1088/0004-637X/745/2/193}{{\em Astrophys. J.} {\bf
  745} (2012)  193}, \href{http://arxiv.org/abs/1105.3516}{{\tt arXiv:1105.3516
  [astro-ph.HE]}}.

\bibitem{Agostini:2019yuq}
{\bf Borexino} Collaboration, M.~Agostini {\em et al.}, ``{Search for
  low-energy neutrinos from astrophysical sources with Borexino},''
  \href{http://dx.doi.org/10.1016/j.astropartphys.2020.102509}{{\em Astropart.
  Phys.} {\bf 125} (2021)  102509}, \href{http://arxiv.org/abs/1909.02422}{{\tt
  arXiv:1909.02422 [hep-ex]}}.

\bibitem{Dasgupta:2019cae}
B.~Dasgupta, R.~Laha, and A.~Ray, ``{Neutrino and positron constraints on
  spinning primordial black hole dark matter},''
  \href{http://dx.doi.org/10.1103/PhysRevLett.125.101101}{{\em Phys. Rev.
  Lett.} {\bf 125} (2020) no.~10, 101101},
  \href{http://arxiv.org/abs/1912.01014}{{\tt arXiv:1912.01014 [hep-ph]}}.

\bibitem{Laha:2020ivk}
R.~Laha, J.~B. Mu\~noz, and T.~R. Slatyer, ``{INTEGRAL constraints on
  primordial black holes and particle dark matter},''
  \href{http://dx.doi.org/10.1103/PhysRevD.101.123514}{{\em Phys. Rev. D} {\bf
  101} (2020) no.~12, 123514}, \href{http://arxiv.org/abs/2004.00627}{{\tt
  arXiv:2004.00627 [astro-ph.CO]}}.

\bibitem{Acharya:2020jbv}
S.~K. Acharya and R.~Khatri, ``{CMB and BBN constraints on evaporating
  primordial black holes revisited},''
  \href{http://dx.doi.org/10.1088/1475-7516/2020/06/018}{{\em JCAP} {\bf 06}
  (2020)  018}, \href{http://arxiv.org/abs/2002.00898}{{\tt arXiv:2002.00898
  [astro-ph.CO]}}.

\bibitem{Smyth:2019whb}
N.~Smyth, S.~Profumo, S.~English, T.~Jeltema, K.~McKinnon, and P.~Guhathakurta,
  ``{Updated Constraints on Asteroid-Mass Primordial Black Holes as Dark
  Matter},'' \href{http://dx.doi.org/10.1103/PhysRevD.101.063005}{{\em Phys.
  Rev. D} {\bf 101} (2020) no.~6, 063005},
  \href{http://arxiv.org/abs/1910.01285}{{\tt arXiv:1910.01285 [astro-ph.CO]}}.

\bibitem{Tisserand:2006zx}
{\bf EROS-2} Collaboration, P.~Tisserand {\em et al.}, ``{Limits on the Macho
  Content of the Galactic Halo from the EROS-2 Survey of the Magellanic
  Clouds},'' \href{http://dx.doi.org/10.1051/0004-6361:20066017}{{\em Astron.
  Astrophys.} {\bf 469} (2007)  387--404},
  \href{http://arxiv.org/abs/astro-ph/0607207}{{\tt arXiv:astro-ph/0607207}}.

\bibitem{Niikura:2019kqi}
H.~Niikura, M.~Takada, S.~Yokoyama, T.~Sumi, and S.~Masaki, ``{Constraints on
  Earth-mass primordial black holes from OGLE 5-year microlensing events},''
  \href{http://dx.doi.org/10.1103/PhysRevD.99.083503}{{\em Phys. Rev. D} {\bf
  99} (2019) no.~8, 083503}, \href{http://arxiv.org/abs/1901.07120}{{\tt
  arXiv:1901.07120 [astro-ph.CO]}}.

\bibitem{Oguri:2017ock}
M.~Oguri, J.~M. Diego, N.~Kaiser, P.~L. Kelly, and T.~Broadhurst,
  ``{Understanding caustic crossings in giant arcs: characteristic scales,
  event rates, and constraints on compact dark matter},''
  \href{http://dx.doi.org/10.1103/PhysRevD.97.023518}{{\em Phys. Rev. D} {\bf
  97} (2018) no.~2, 023518}, \href{http://arxiv.org/abs/1710.00148}{{\tt
  arXiv:1710.00148 [astro-ph.CO]}}.

\bibitem{Ali-Haimoud:2017rtz}
Y.~Ali-Haïmoud, E.~D. Kovetz, and M.~Kamionkowski, ``{Merger rate of
  primordial black-hole binaries},''
  \href{http://dx.doi.org/10.1103/PhysRevD.96.123523}{{\em Phys. Rev.} {\bf
  D96} (2017) no.~12, 123523},
\href{http://arxiv.org/abs/1709.06576}{{\tt arXiv:1709.06576 [astro-ph.CO]}}.

\bibitem{Bird:2016dcv}
S.~Bird, I.~Cholis, J.~B. Mu\~noz, Y.~Ali-Haïmoud, M.~Kamionkowski, E.~D.
  Kovetz, A.~Raccanelli, and A.~G. Riess, ``{Did LIGO detect dark matter?},''
  \href{http://dx.doi.org/10.1103/PhysRevLett.116.201301}{{\em Phys. Rev.
  Lett.} {\bf 116} (2016) no.~20, 201301},
  \href{http://arxiv.org/abs/1603.00464}{{\tt arXiv:1603.00464 [astro-ph.CO]}}.

\bibitem{Sasaki:2016jop}
M.~Sasaki, T.~Suyama, T.~Tanaka, and S.~Yokoyama, ``{Primordial Black Hole
  Scenario for the Gravitational-Wave Event GW150914},''
  \href{http://dx.doi.org/10.1103/PhysRevLett.117.061101}{{\em Phys. Rev.
  Lett.} {\bf 117} (2016) no.~6, 061101},
  \href{http://arxiv.org/abs/1603.08338}{{\tt arXiv:1603.08338 [astro-ph.CO]}}.
  [Erratum: Phys.Rev.Lett. 121, 059901 (2018)].

\bibitem{Cholis:2016kqi}
I.~Cholis, E.~D. Kovetz, Y.~Ali-Haïmoud, S.~Bird, M.~Kamionkowski, J.~B.
  Mu\~noz, and A.~Raccanelli, ``{Orbital eccentricities in primordial black
  hole binaries},'' \href{http://dx.doi.org/10.1103/PhysRevD.94.084013}{{\em
  Phys. Rev. D} {\bf 94} (2016) no.~8, 084013},
  \href{http://arxiv.org/abs/1606.07437}{{\tt arXiv:1606.07437 [astro-ph.HE]}}.

\bibitem{Clesse:2016vqa}
S.~Clesse and J.~Garc\'\i{}a-Bellido, ``{The clustering of massive Primordial
  Black Holes as Dark Matter: measuring their mass distribution with Advanced
  LIGO},'' \href{http://dx.doi.org/10.1016/j.dark.2016.10.002}{{\em Phys. Dark
  Univ.} {\bf 15} (2017)  142--147},
  \href{http://arxiv.org/abs/1603.05234}{{\tt arXiv:1603.05234 [astro-ph.CO]}}.

\bibitem{Raccanelli:2016cud}
A.~Raccanelli, E.~D. Kovetz, S.~Bird, I.~Cholis, and J.~B. Munoz,
  ``{Determining the progenitors of merging black-hole binaries},''
  \href{http://dx.doi.org/10.1103/PhysRevD.94.023516}{{\em Phys. Rev. D} {\bf
  94} (2016) no.~2, 023516}, \href{http://arxiv.org/abs/1605.01405}{{\tt
  arXiv:1605.01405 [astro-ph.CO]}}.

\bibitem{Kovetz:2017rvv}
E.~D. Kovetz, ``{Probing Primordial-Black-Hole Dark Matter with Gravitational
  Waves},'' \href{http://dx.doi.org/10.1103/PhysRevLett.119.131301}{{\em Phys.
  Rev. Lett.} {\bf 119} (2017) no.~13, 131301},
  \href{http://arxiv.org/abs/1705.09182}{{\tt arXiv:1705.09182 [astro-ph.CO]}}.

\bibitem{Authors:2019qbw}
{\bf LIGO Scientific, Virgo} Collaboration, B.~Abbott {\em et al.}, ``{Search
  for Subsolar Mass Ultracompact Binaries in Advanced LIGO\textquoteright{}s
  Second Observing Run},''
  \href{http://dx.doi.org/10.1103/PhysRevLett.123.161102}{{\em Phys. Rev.
  Lett.} {\bf 123} (2019) no.~16, 161102},
  \href{http://arxiv.org/abs/1904.08976}{{\tt arXiv:1904.08976 [astro-ph.CO]}}.

\bibitem{Kavanagh:2018ggo}
B.~J. Kavanagh, D.~Gaggero, and G.~Bertone, ``{Merger rate of a subdominant
  population of primordial black holes},''
  \href{http://dx.doi.org/10.1103/PhysRevD.98.023536}{{\em Phys. Rev. D} {\bf
  98} (2018) no.~2, 023536}, \href{http://arxiv.org/abs/1805.09034}{{\tt
  arXiv:1805.09034 [astro-ph.CO]}}.

\bibitem{DeLuca:2020qqa}
V.~De~Luca, G.~Franciolini, P.~Pani, and A.~Riotto, ``{Primordial Black Holes
  Confront LIGO/Virgo data: Current situation},''
  \href{http://dx.doi.org/10.1088/1475-7516/2020/06/044}{{\em JCAP} {\bf 06}
  (2020)  044}, \href{http://arxiv.org/abs/2005.05641}{{\tt arXiv:2005.05641
  [astro-ph.CO]}}.

\bibitem{Wang:2016ana}
S.~Wang, Y.-F. Wang, Q.-G. Huang, and T.~G.~F. Li, ``{Constraints on the
  Primordial Black Hole Abundance from the First Advanced LIGO Observation Run
  Using the Stochastic Gravitational-Wave Background},''
  \href{http://dx.doi.org/10.1103/PhysRevLett.120.191102}{{\em Phys. Rev.
  Lett.} {\bf 120} (2018) no.~19, 191102},
  \href{http://arxiv.org/abs/1610.08725}{{\tt arXiv:1610.08725 [astro-ph.CO]}}.

\bibitem{carr1981pregalactic}
B.~Carr, ``Pregalactic black hole accretion and the thermal history of the
  universe,'' {\em Monthly Notices of the Royal Astronomical Society} {\bf 194}
  (1981) no.~3, 639--668.

\bibitem{Ricotti:2007au}
M.~Ricotti, J.~P. Ostriker, and K.~J. Mack, ``{Effect of Primordial Black Holes
  on the Cosmic Microwave Background and Cosmological Parameter Estimates},''
  \href{http://dx.doi.org/10.1086/587831}{{\em Astrophys. J.} {\bf 680} (2008)
  829}, \href{http://arxiv.org/abs/0709.0524}{{\tt arXiv:0709.0524
  [astro-ph]}}.

\bibitem{Serpico:2020ehh}
P.~D. Serpico, V.~Poulin, D.~Inman, and K.~Kohri, ``{Cosmic microwave
  background bounds on primordial black holes including dark matter halo
  accretion},'' \href{http://dx.doi.org/10.1103/PhysRevResearch.2.023204}{{\em
  Phys. Rev. Res.} {\bf 2} (2020) no.~2, 023204},
  \href{http://arxiv.org/abs/2002.10771}{{\tt arXiv:2002.10771 [astro-ph.CO]}}.

\bibitem{Carr:2020gox}
B.~Carr, K.~Kohri, Y.~Sendouda, and J.~Yokoyama, ``{Constraints on Primordial
  Black Holes},'' \href{http://arxiv.org/abs/2002.12778}{{\tt arXiv:2002.12778
  [astro-ph.CO]}}.

\bibitem{Green:2020jor}
A.~M. Green and B.~J. Kavanagh, ``{Primordial Black Holes as a dark matter
  candidate},'' \href{http://dx.doi.org/10.1088/1361-6471/abc534}{{\em J. Phys.
  G} {\bf 48} (2021) no.~4, 043001},
  \href{http://arxiv.org/abs/2007.10722}{{\tt arXiv:2007.10722 [astro-ph.CO]}}.

\bibitem{Khlopov:2008qy}
M.~Y. Khlopov, ``{Primordial Black Holes},''
  \href{http://dx.doi.org/10.1088/1674-4527/10/6/001}{{\em Res. Astron.
  Astrophys.} {\bf 10} (2010)  495--528},
  \href{http://arxiv.org/abs/0801.0116}{{\tt arXiv:0801.0116 [astro-ph]}}.

\bibitem{Espinosa:2018eve}
J.~R. Espinosa, D.~Racco, and A.~Riotto, ``{A Cosmological Signature of the SM
  Higgs Instability: Gravitational Waves},''
  \href{http://dx.doi.org/10.1088/1475-7516/2018/09/012}{{\em JCAP} {\bf 09}
  (2018)  012}, \href{http://arxiv.org/abs/1804.07732}{{\tt arXiv:1804.07732
  [hep-ph]}}.

\bibitem{Kohri:2018awv}
K.~Kohri and T.~Terada, ``{Semianalytic calculation of gravitational wave
  spectrum nonlinearly induced from primordial curvature perturbations},''
  \href{http://dx.doi.org/10.1103/PhysRevD.97.123532}{{\em Phys. Rev. D} {\bf
  97} (2018) no.~12, 123532}, \href{http://arxiv.org/abs/1804.08577}{{\tt
  arXiv:1804.08577 [gr-qc]}}.

\bibitem{Aggarwal:2018mgp}
K.~Aggarwal {\em et al.}, ``{The NANOGrav 11-Year Data Set: Limits on
  Gravitational Waves from Individual Supermassive Black Hole Binaries},''
  \href{http://dx.doi.org/10.3847/1538-4357/ab2236}{{\em Astrophys. J.} {\bf
  880} (2019)  2}, \href{http://arxiv.org/abs/1812.11585}{{\tt arXiv:1812.11585
  [astro-ph.GA]}}.

\bibitem{NANOGrav:2020bcs}
{\bf NANOGrav} Collaboration, Z.~Arzoumanian {\em et al.}, ``{The NANOGrav 12.5
  yr Data Set: Search for an Isotropic Stochastic Gravitational-wave
  Background},'' \href{http://dx.doi.org/10.3847/2041-8213/abd401}{{\em
  Astrophys. J. Lett.} {\bf 905} (2020) no.~2, L34},
  \href{http://arxiv.org/abs/2009.04496}{{\tt arXiv:2009.04496 [astro-ph.HE]}}.

\bibitem{Lentati:2015qwp}
L.~Lentati {\em et al.}, ``{European Pulsar Timing Array Limits On An Isotropic
  Stochastic Gravitational-Wave Background},''
  \href{http://dx.doi.org/10.1093/mnras/stv1538}{{\em Mon. Not. Roy. Astron.
  Soc.} {\bf 453} (2015) no.~3, 2576--2598},
  \href{http://arxiv.org/abs/1504.03692}{{\tt arXiv:1504.03692 [astro-ph.CO]}}.

\bibitem{Shannon:2015ect}
R.~M. Shannon {\em et al.}, ``{Gravitational waves from binary supermassive
  black holes missing in pulsar observations},''
  \href{http://dx.doi.org/10.1126/science.aab1910}{{\em Science} {\bf 349}
  (2015) no.~6255, 1522--1525}, \href{http://arxiv.org/abs/1509.07320}{{\tt
  arXiv:1509.07320 [astro-ph.CO]}}.

\bibitem{Qin:2018yhy}
W.~Qin, K.~K. Boddy, M.~Kamionkowski, and L.~Dai, ``{Pulsar-timing arrays,
  astrometry, and gravitational waves},''
  \href{http://dx.doi.org/10.1103/PhysRevD.99.063002}{{\em Phys. Rev. D} {\bf
  99} (2019) no.~6, 063002}, \href{http://arxiv.org/abs/1810.02369}{{\tt
  arXiv:1810.02369 [astro-ph.CO]}}.

\bibitem{LIGOScientific:2019vic}
{\bf LIGO Scientific, Virgo} Collaboration, B.~P. Abbott {\em et al.},
  ``{Search for the isotropic stochastic background using data from Advanced
  LIGO\textquoteright{}s second observing run},''
  \href{http://dx.doi.org/10.1103/PhysRevD.100.061101}{{\em Phys. Rev. D} {\bf
  100} (2019) no.~6, 061101}, \href{http://arxiv.org/abs/1903.02886}{{\tt
  arXiv:1903.02886 [gr-qc]}}.

\bibitem{LIGOScientific:2016aoc}
{\bf LIGO Scientific, Virgo} Collaboration, B.~P. Abbott {\em et al.},
  ``{Observation of Gravitational Waves from a Binary Black Hole Merger},''
  \href{http://dx.doi.org/10.1103/PhysRevLett.116.061102}{{\em Phys. Rev.
  Lett.} {\bf 116} (2016) no.~6, 061102},
  \href{http://arxiv.org/abs/1602.03837}{{\tt arXiv:1602.03837 [gr-qc]}}.

\bibitem{LIGOScientific:2016dsl}
{\bf LIGO Scientific, Virgo} Collaboration, B.~P. Abbott {\em et al.},
  ``{Binary Black Hole Mergers in the first Advanced LIGO Observing Run},''
  \href{http://dx.doi.org/10.1103/PhysRevX.6.041015}{{\em Phys. Rev. X} {\bf 6}
  (2016) no.~4, 041015}, \href{http://arxiv.org/abs/1606.04856}{{\tt
  arXiv:1606.04856 [gr-qc]}}. [Erratum: Phys.Rev.X 8, 039903 (2018)].

\bibitem{LIGOScientific:2017ycc}
{\bf LIGO Scientific, Virgo} Collaboration, B.~P. Abbott {\em et al.},
  ``{GW170814: A Three-Detector Observation of Gravitational Waves from a
  Binary Black Hole Coalescence},''
  \href{http://dx.doi.org/10.1103/PhysRevLett.119.141101}{{\em Phys. Rev.
  Lett.} {\bf 119} (2017) no.~14, 141101},
  \href{http://arxiv.org/abs/1709.09660}{{\tt arXiv:1709.09660 [gr-qc]}}.

\bibitem{Akutsu:2015hua}
{\bf KAGRA} Collaboration, T.~Akutsu, ``{Large-scale cryogenic
  gravitational-wave telescope in Japan: KAGRA},''
  \href{http://dx.doi.org/10.1088/1742-6596/610/1/012016}{{\em J. Phys. Conf.
  Ser.} {\bf 610} (2015) no.~1, 012016}.

\bibitem{Haino:2020age}
{\bf KAGRA} Collaboration, S.~Haino, {\em {KAGRA, Underground Cryogenic
  Gravitational Wave Telescope}},
  \href{http://dx.doi.org/10.1142/9789811207402_0012}{pp.~174--184}.
\newblock WSP, Singapur, 2020.

\bibitem{Maggiore:2019uih}
M.~Maggiore {\em et al.}, ``{Science Case for the Einstein Telescope},''
  \href{http://dx.doi.org/10.1088/1475-7516/2020/03/050}{{\em JCAP} {\bf 03}
  (2020)  050}, \href{http://arxiv.org/abs/1912.02622}{{\tt arXiv:1912.02622
  [astro-ph.CO]}}.

\bibitem{LISA:2017pwj}
{\bf LISA} Collaboration, P.~Amaro-Seoane {\em et al.}, ``{Laser Interferometer
  Space Antenna},'' \href{http://arxiv.org/abs/1702.00786}{{\tt
  arXiv:1702.00786 [astro-ph.IM]}}.

\bibitem{Kaiser:2020tlg}
A.~R. Kaiser and S.~T. McWilliams, ``{Sensitivity of present and future
  detectors across the black-hole binary gravitational wave spectrum},''
  \href{http://dx.doi.org/10.1088/1361-6382/abd4f6}{{\em Class. Quant. Grav.}
  {\bf 38} (2021) no.~5, 055009}, \href{http://arxiv.org/abs/2010.02135}{{\tt
  arXiv:2010.02135 [gr-qc]}}.

\bibitem{Barausse:2020rsu}
E.~Barausse {\em et al.}, ``{Prospects for Fundamental Physics with LISA},''
  \href{http://dx.doi.org/10.1007/s10714-020-02691-1}{{\em Gen. Rel. Grav.}
  {\bf 52} (2020) no.~8, 81}, \href{http://arxiv.org/abs/2001.09793}{{\tt
  arXiv:2001.09793 [gr-qc]}}.

\bibitem{LISACosmologyWorkingGroup:2022kbp}
{\bf LISA Cosmology Working Group} Collaboration, N.~Bartolo {\em et al.},
  ``{Probing Anisotropies of the Stochastic Gravitational Wave Background with
  LISA},'' \href{http://arxiv.org/abs/2201.08782}{{\tt arXiv:2201.08782
  [astro-ph.CO]}}.

\bibitem{Seto:2001qf}
N.~Seto, S.~Kawamura, and T.~Nakamura, ``{Possibility of direct measurement of
  the acceleration of the universe using 0.1-Hz band laser interferometer
  gravitational wave antenna in space},''
  \href{http://dx.doi.org/10.1103/PhysRevLett.87.221103}{{\em Phys. Rev. Lett.}
  {\bf 87} (2001)  221103}, \href{http://arxiv.org/abs/astro-ph/0108011}{{\tt
  arXiv:astro-ph/0108011}}.

\bibitem{Yagi:2011wg}
K.~Yagi and N.~Seto, ``{Detector configuration of DECIGO/BBO and identification
  of cosmological neutron-star binaries},''
  \href{http://dx.doi.org/10.1103/PhysRevD.83.044011}{{\em Phys. Rev. D} {\bf
  83} (2011)  044011}, \href{http://arxiv.org/abs/1101.3940}{{\tt
  arXiv:1101.3940 [astro-ph.CO]}}. [Erratum: Phys.Rev.D 95, 109901 (2017)].

\bibitem{Kawamura:2020pcg}
S.~Kawamura {\em et al.}, ``{Current status of space gravitational wave antenna
  DECIGO and B-DECIGO},'' \href{http://dx.doi.org/10.1093/ptep/ptab019}{{\em
  PTEP} {\bf 2021} (2021) no.~5, 05A105},
  \href{http://arxiv.org/abs/2006.13545}{{\tt arXiv:2006.13545 [gr-qc]}}.

\bibitem{Badurina:2019hst}
L.~Badurina {\em et al.}, ``{AION: An Atom Interferometer Observatory and
  Network},'' \href{http://dx.doi.org/10.1088/1475-7516/2020/05/011}{{\em JCAP}
  {\bf 05} (2020)  011}, \href{http://arxiv.org/abs/1911.11755}{{\tt
  arXiv:1911.11755 [astro-ph.CO]}}.

\bibitem{Ruan:2018tsw}
W.-H. Ruan, Z.-K. Guo, R.-G. Cai, and Y.-Z. Zhang, ``{Taiji program:
  Gravitational-wave sources},''
  \href{http://dx.doi.org/10.1142/S0217751X2050075X}{{\em Int. J. Mod. Phys. A}
  {\bf 35} (2020) no.~17, 2050075}, \href{http://arxiv.org/abs/1807.09495}{{\tt
  arXiv:1807.09495 [gr-qc]}}.

\bibitem{TianQin:2015yph}
{\bf TianQin} Collaboration, J.~Luo {\em et al.}, ``{TianQin: a space-borne
  gravitational wave detector},''
  \href{http://dx.doi.org/10.1088/0264-9381/33/3/035010}{{\em Class. Quant.
  Grav.} {\bf 33} (2016) no.~3, 035010},
  \href{http://arxiv.org/abs/1512.02076}{{\tt arXiv:1512.02076 [astro-ph.IM]}}.

\bibitem{Carr:2009jm}
B.~Carr, K.~Kohri, Y.~Sendouda, and J.~Yokoyama, ``{New cosmological
  constraints on primordial black holes},''
  \href{http://dx.doi.org/10.1103/PhysRevD.81.104019}{{\em Phys. Rev. D} {\bf
  81} (2010)  104019}, \href{http://arxiv.org/abs/0912.5297}{{\tt
  arXiv:0912.5297 [astro-ph.CO]}}.

\bibitem{Chluba:2020oip}
J.~Chluba, A.~Ravenni, and S.~K. Acharya, ``{Thermalization of large energy
  release in the early Universe},''
  \href{http://dx.doi.org/10.1093/mnras/staa2131}{{\em Mon. Not. Roy. Astron.
  Soc.} {\bf 498} (2020) no.~1, 959--980},
  \href{http://arxiv.org/abs/2005.11325}{{\tt arXiv:2005.11325 [astro-ph.CO]}}.

\bibitem{Carr:2016hva}
B.~J. Carr, K.~Kohri, Y.~Sendouda, and J.~Yokoyama, ``{Constraints on
  primordial black holes from the Galactic gamma-ray background},''
  \href{http://dx.doi.org/10.1103/PhysRevD.94.044029}{{\em Phys. Rev. D} {\bf
  94} (2016) no.~4, 044029}, \href{http://arxiv.org/abs/1604.05349}{{\tt
  arXiv:1604.05349 [astro-ph.CO]}}.

\bibitem{Boudaud:2018hqb}
M.~Boudaud and M.~Cirelli, ``{Voyager 1 $e^\pm$ Further Constrain Primordial
  Black Holes as Dark Matter},''
  \href{http://dx.doi.org/10.1103/PhysRevLett.122.041104}{{\em Phys. Rev.
  Lett.} {\bf 122} (2019) no.~4, 041104},
  \href{http://arxiv.org/abs/1807.03075}{{\tt arXiv:1807.03075 [astro-ph.HE]}}.

\bibitem{Papanikolaou:2020qtd}
T.~Papanikolaou, V.~Vennin, and D.~Langlois, ``{Gravitational waves from a
  universe filled with primordial black holes},''
  \href{http://dx.doi.org/10.1088/1475-7516/2021/03/053}{{\em JCAP} {\bf 03}
  (2021)  053}, \href{http://arxiv.org/abs/2010.11573}{{\tt arXiv:2010.11573
  [astro-ph.CO]}}.

\bibitem{Papanikolaou:2022chm}
T.~Papanikolaou, ``{Gravitational waves induced from primordial black hole
  fluctuations: the~effect of an extended mass function},''
  \href{http://dx.doi.org/10.1088/1475-7516/2022/10/089}{{\em JCAP} {\bf 10}
  (2022)  089}, \href{http://arxiv.org/abs/2207.11041}{{\tt arXiv:2207.11041
  [astro-ph.CO]}}.

\bibitem{Ireland:2023avg}
A.~Ireland, S.~Profumo, and J.~Scharnhorst, ``{Primordial Gravitational Waves
  From Black Hole Evaporation in Standard and Non-Standard Cosmologies},''
  \href{http://arxiv.org/abs/2302.10188}{{\tt arXiv:2302.10188 [gr-qc]}}.

\bibitem{Allahverdi:2020bys}
R.~Allahverdi {\em et al.}, ``{The First Three Seconds: a Review of Possible
  Expansion Histories of the Early Universe},''
  \href{http://arxiv.org/abs/2006.16182}{{\tt arXiv:2006.16182 [astro-ph.CO]}}.

\bibitem{Domenech:2019quo}
G.~Dom\`enech, ``{Induced gravitational waves in a general cosmological
  background},'' \href{http://dx.doi.org/10.1142/S0218271820500285}{{\em Int.
  J. Mod. Phys. D} {\bf 29} (2020) no.~03, 2050028},
  \href{http://arxiv.org/abs/1912.05583}{{\tt arXiv:1912.05583 [gr-qc]}}.

\bibitem{Bhattacharya:2019bvk}
S.~Bhattacharya, S.~Mohanty, and P.~Parashari, ``{Primordial black holes and
  gravitational waves in nonstandard cosmologies},''
  \href{http://dx.doi.org/10.1103/PhysRevD.102.043522}{{\em Phys. Rev. D} {\bf
  102} (2020) no.~4, 043522}, \href{http://arxiv.org/abs/1912.01653}{{\tt
  arXiv:1912.01653 [astro-ph.CO]}}.

\bibitem{Caprini:2019pxz}
C.~Caprini, D.~G. Figueroa, R.~Flauger, G.~Nardini, M.~Peloso, M.~Pieroni,
  A.~Ricciardone, and G.~Tasinato, ``{Reconstructing the spectral shape of a
  stochastic gravitational wave background with LISA},''
  \href{http://dx.doi.org/10.1088/1475-7516/2019/11/017}{{\em JCAP} {\bf 11}
  (2019)  017}, \href{http://arxiv.org/abs/1906.09244}{{\tt arXiv:1906.09244
  [astro-ph.CO]}}.

\bibitem{2018cosp...42E1401H}
M.~{Hazumi}, ``{LiteBIRD for B-mode from space},'' in {\em 42nd COSPAR
  Scientific Assembly}, vol.~42, pp.~E1.2--20--18.
\newblock July, 2018.

\bibitem{CMB-S4:2016ple}
{\bf CMB-S4} Collaboration, K.~N. Abazajian {\em et al.}, ``{CMB-S4 Science
  Book, First Edition},'' \href{http://arxiv.org/abs/1610.02743}{{\tt
  arXiv:1610.02743 [astro-ph.CO]}}.

\end{thebibliography}\endgroup

\end{document}